\documentclass[letter,aps,prd,10pt,preprintnumbers,superscriptaddress,nofootinbib,amsmath,amssymb,floatfix
]{revtex4}
\usepackage{times}
\usepackage{hyperref}
\usepackage{amsmath}
\usepackage{amssymb}
\usepackage{graphicx}
\usepackage{subfigure}
\usepackage{booktabs}
\usepackage{rotating}
\usepackage{longtable}
\usepackage{bm}
\usepackage{float}
\usepackage{epstopdf}
\usepackage{xcolor}
\usepackage{comment}

\makeatletter

\newcommand{\Rmnum}[1]{\expandafter\@slowromancap\romannumeral #1@}
\makeatother
\begin{document}
\makeatletter
\newcommand*{\rom}[1]{\expandafter\@slowromancap\romannumeral #1@}
\makeatother

\title{\boldmath Astrophysical Signature and Optical Appearance of Weyl-Corrected Einstein-Maxwell Black Holes}
\author{Hassan Hassanabadi}
\email{hassanhassanabadi@mail.fresnostate.edu}
\affiliation{Physics Department, California State University, Fresno, CA 93740, USA}
\affiliation{Department   of   Physics, Faculty of Science,   University   of   Hradec   Kr\'{a}lov\'{e},  Rokitansk\'{e}ho 62, 500   03   Hradec   Kr\'{a}lov\'{e},   Czechia}
\affiliation{Khazar University, Department of Physics and Electronics, 41 Mahsati Str, AZ1096, Baku, Azerbaijan}
\author{Mrinnoy M. Gohain}
\email{mrinmoygohain19@gmail.com}
\affiliation{Department of Physics, Dibrugarh University, Dibrugarh Assam, India, 786004}
\author{Kalyan Bhuyan}
\email{kalyanbhuyan@dibru.ac.in}
\affiliation{Department of Physics, Dibrugarh University, Dibrugarh Assam, India, 786004}
\affiliation{Theoretical Physics Division, Centre for Atmospheric Studies, Dibrugarh University, Dibrugarh, Assam, India 786004}
\author{Farokhnaz Hosseinifar}
\email{f.hoseinifar94@gmail.com}
\affiliation{Center for Theoretical Physics, Khazar University, 41 Mehseti Street, Baku, AZ-1096, Azerbaijan}

\begin{abstract}
	\begin{center}
	\textbf{Abstract}
	\end{center}
	In this work, we investigate the physics of charged black holes modified by Weyl corrections, which emerge from the non-minimal coupling between spacetime curvature and electromagnetism. We begin by revisiting the thermodynamics of these systems, deriving the Hawking temperature, Wald entropy, and heat capacity to examine how the Weyl correction parameter reshapes the landscape of thermal stability and phase transitions.
	Then, we apply a topological method to classify the thermodynamic states and reveal the impact of the Weyl modifications on this classification.
		To explore astrophysical signatures, we analyze the null trajectories and shadow cast by photons under two-photon polarization modes, deriving observational constraints on the black hole parameters.
		Finally, we model the accretion disk around these black holes. By calculating the energy flux, spectral luminosity, and differential luminosity, we show how these corrections leave a detectable trace on the emitted radiation.
	\end{abstract}
\maketitle
	\section{Introduction}\label{Sec1}
The prediction of black holes remains one of the most profound triumphs of Albert Einstein’s general relativity \cite{einstein1916foundation,schwarzschild1916gravitationsfeld,oppenheimer1939continued,chow2008gravity,yagi2016black}.
While the classical Schwarzschild and Kerr metrics provide a robust foundation for understanding these objects, the quest for a more complete description of spacetime has led physicists to look beyond the standard framework \cite{boi2004theories,hawking2010nature,blum2015reinvention,witten2022note}. The motivation for exploring generalized and modified gravity models stems from the inherent limitations of general relativity, specifically the presence of singularities and the ongoing challenge of reconciling gravity with quantum mechanics \cite{dewitt1967quantum,hawking1970singularities,stelle1977renormalization,santilli1995limitations,donoghue1994general,riess1998observational}. Consequently, a diverse array of modified models has emerged. These include, for instance, regular black holes (such as the Bardeen \cite{bardeen1968non} or Hayward \cite{hayward2006formation} metrics) designed to bypass the singularity problem \cite{ayon1998regular,ansoldi2008spherical,rodrigues2016generalisation}, f(R) gravity \cite{starobinsky1980new,capozziello2002curvature,sotiriou2010f}, which modifies the Einstein-Hilbert action \cite{stelle1977renormalization,lovelock1971einstein} to explain cosmological acceleration, and black holes in the presence of non-linear electrodynamics like the Born-Infeld theory \cite{born1934foundations,gunasekaran2012extended}.

The introduction of these modified solutions has generated a vast volume of literature dedicated to probing their observational signatures. Extensive research has been conducted on the thermodynamic stability and phase transitions of these new black holes, revealing how corrections to the action affect their Hawking temperature and entropy \cite{hawking1983thermodynamics,kastor2009enthalpy,kubizvnak2012p,mahapatra2016thermodynamics,dey2018holographic,zhang2025new,ma2024effectiveness}. Similarly, the study of black hole shadows has become a pivotal tool for testing these models against data from the Event Horizon Telescope \cite{falcke1999viewing,johannsen2010testing,huang2016double,mureika2017black,event2019first,gralla2019black,ali2019shadow,fathi2021ergosphere,chen2024kerr}, while investigations into accretion disks and gravitational lensing provide further insight into how modified gravity alters the path of light and matter in the vicinity of the horizon \cite{novikov1973astrophysics,page1974disk,shakura1973black,bambi2012code,abramowicz2013foundations,nieto2025accretion}.
A significant amount of attention has been devoted to the quasinormal modes of these spacetimes \cite{kokkotas1999quasi,cardoso2001quasinormal,cardoso2003quasinormal,konoplya2003quasinormal,cardoso2004quasinormal,berti2006gravitational,berti2009quasinormal,konoplya2011quasinormal,zinhailo2018quasinormal,momennia2019near,churilova2019analytical,momennia2020quasinormal,konoplya2021conformal,konoplya2023quasinormal,hosseinifar2025quasinormal,konoplya2025quasinormal,hassanabadi2026periodic}.
Furthermore, the topological nature of black holes has recently emerged as a powerful tool for classifying these solutions. By employing methods such as vector field analysis, researchers have investigated the topological charge of these modified black holes. This method enhances our understanding of thermodynamic phase transitions and allows us to view black holes as topological defects in the thermodynamic parameter space \cite{wei2019intrinsic,wei2022topology,wei2022black,wei2023topology,wei2024universal,dong2025thermodynamic,zare2025accretion,wu2025extended}.

Among the most compelling generalizations is the Einstein-Maxwell theory \cite{reissner1916eigengravitation,nordstrom1918energy}, which describes the intricate coupling between the gravitational metric and the electromagnetic field. This framework allows for the existence of charged black holes, where the energy-momentum tensor of the Maxwell field acts as a source for spacetime curvature \cite{misner1973gravitation}. In its generalized form, this theory incorporates non-minimal couplings and higher-order terms \cite{horndeski1976conservation}, offering a bridge between classical gravity and the corrections expected from effective field theories of quantum gravity \cite{drummond1980qed,shore2003quantum,balakin2005non}.

Within the landscape of non-minimally coupled theories, the introduction of Weyl corrections represents a significant step toward capturing the subtle interplay between the gravitational field and electromagnetic phenomena. These corrections involve a specific coupling between the Weyl tensor, which represents the free-space part of the gravitational field \cite{penrose1965zero}, and the Maxwell field strength. Such interactions are not merely mathematical extensions but are deeply rooted in the effective actions of quantum electrodynamics in curved backgrounds, reflecting the vacuum polarization effects that arise in regions of extreme spacetime curvature \cite{drummond1980qed,shore2007superluminality}. By breaking the conformal invariance in Maxwell’s action \cite{horndeski1976conservation}, one carves out a richer playground for studying charged black holes outside classical boundaries.

Extensive research has been conducted on the physical effects that Weyl-corrected black holes produce in different scientific areas. Significant effort has been devoted to studying the thermodynamic properties of these systems by calculating their modified Hawking temperature, entropy, and heat capacity to understand how the coupling parameter affects system stability and phase transitions in Refs. \cite{dey2015thermodynamics,mahapatra2016thermodynamics,sharif2020quasi}.
Ref. \cite{al2025static} has developed the effective potential and Innermost Stable Circular Orbit (ISCO), and black hole shadow description through particle trajectory analysis in these background environments. Moreover, the scope of this research has been broadened to include rotating solutions, which show how Weyl coupling affects black holes with angular momentum by creating complex changes in their event horizon and ergosphere structures \cite{chen2014rotating,ali2019shadow}.

To complement these studies, this work investigates static charged Weyl-corrected black holes by systematically connecting their fundamental physics to observable features. We focus on quantifying the influence of the coupling parameter across thermodynamic stability, polarization-dependent photon dynamics, and accretion disk signatures.
The paper is organized as follows: Section \ref{Sec4} presents an analysis of thermodynamic properties for the Weyl-corrected black hole.
We establish the essential thermodynamic values through event horizon analysis, which includes derivation of Wald entropy, Hawking temperature, heat capacity evaluation, and free energy calculation to determine black hole stability. We adopt a modern approach to study the topological classification of the black hole in Section \ref{Sec6} through its thermodynamic potentials. These serve as topological defects that we analyze with the topological charge method. Section \ref{Sec2} presents a study of massless particle movement through geodesic structure analysis, which leads to black hole shadow property identification and finding constraints on the charge and Weyl correction parameter.
Section \ref{sec2.1} investigates the polarization dependence of light trajectories in the Weyl-corrected spacetime.
Section \ref{Sec3} extends our research to study massive particle movement through ISCO derivation and radiative efficiency calculation. We calculate energy flux, spectral luminosity, and differential luminosity to study astrophysical signatures through the accretion process. The final Section \ref{Sec10} presents our main discoveries together with closing statements about the physical meaning of our research outcomes. We work in the units where $G=c=\hbar=1$ and the metric signature is adopted as $(-,\,+,\,+,\,+)$.
\section{Thermodynamic Properties}\label{Sec4}
\begin{figure}[ht]
		 \centerline{
		\includegraphics[width=6.cm]{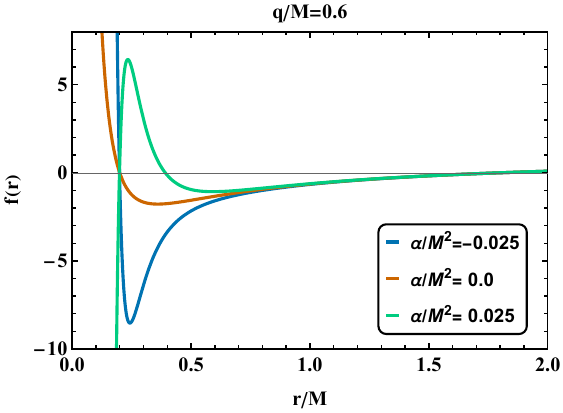} \hspace{0.2cm}
		\includegraphics[width=6.cm]{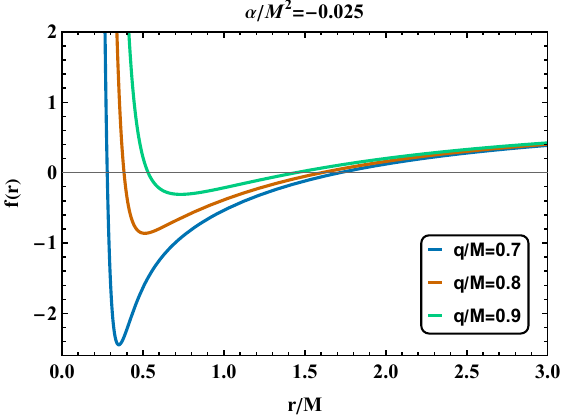} \hspace{0.2cm}
		}
		\centerline{
		\includegraphics[width=6.cm]{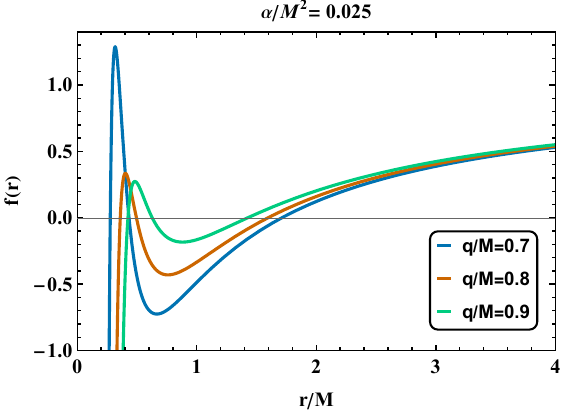} \hspace{-0.2cm}}
	\caption{The behavior of the lapse function as a function of $r/M$. Upper left panel: $q/M=0.6$ for three different values of $\alpha/M^2$. Upper right panel: negative Weyl correction parameter $\alpha/M^2=-0.025$. Lower panel: positive Weyl correction parameter $\alpha/M^2=0.025$.}\label{fig:fr}
\end{figure}

\begin{figure}[ht]
		 \centerline{
		\includegraphics[width=7cm]{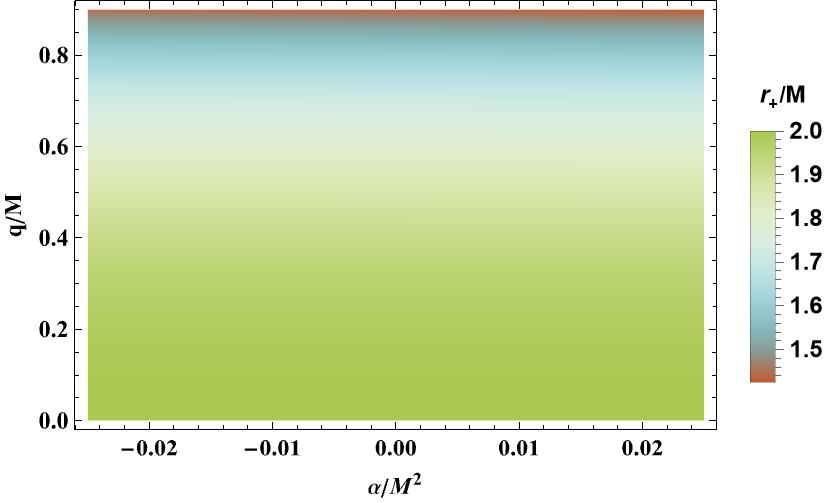} \hspace{-0.2cm}}
	\caption{
		The event horizon radius of the black hole as a function of $\alpha/M^2$ and $q/M$. The magnitude of the horizon radius for each parameter choice is indicated by the color bar.}
	\label{fig:Horizon}
\end{figure}

In this section, we first present the gravitational background of the black hole under consideration. Our analysis is based on a modified gravity model where the electromagnetic sector is non-minimally coupled to the Weyl tensor. Following Ref. \cite{chen2014rotating}, the action is expressed as
\begin{equation}
	S=\int d^4 x\sqrt{-g}\mathcal{L},
\end{equation}
where $g$ is the determinant of the metric tensor and the Lagrangian including a Weyl correction term $\mathcal{L}_{\rm interaction}=\alpha C_{\mu\nu\rho\sigma}F^{\mu\nu}F^{\rho\sigma}$ reads
\begin{equation}
	\mathcal{L}=\frac{1}{16\pi G}R-\frac{1}{4}F_{\mu\nu}F^{\mu\nu}+\alpha C_{\mu\nu\rho\sigma}F^{\mu\nu}F^{\rho\sigma},
\end{equation}
in which $R=g^{\mu\nu}R_{\mu\nu}$ indicates the Ricci scalar, $F_{\mu\nu}=\partial_{\mu}A_{\nu}-\partial_{\nu}A_{\mu}$ represents the electromagnetic field tensor, the parameter $\alpha$ refers to a small coupling with dimensions of [length]$^2$ in $4D$, and $C_{\mu\nu\rho\sigma}$ denotes the Weyl tensor. In the limit of $\alpha\to 0$, the above Lagrangian tends to the standard Einstein-Maxwell Lagrangian.
For $q=0$, the background electromagnetic field vanishes and the background geometry reduces to Schwarzschild. Nevertheless, the Weyl coupling may still modify the propagation of electromagnetic perturbations in this background.
In four dimensions, the Weyl tensor is defined as
\begin{equation}	C_{\mu\nu\rho\sigma}=R_{\mu\nu\rho\sigma}-\frac{1}{2}\big(g_{\mu\rho}R_{\sigma\nu}-g_{\mu\sigma}R_{\rho\nu}-g_{\nu\rho}R_{\sigma\mu}+g_{\nu\sigma}R_{\rho\mu}\big)+\frac{1}{6}R\big(g_{\mu\rho}g_{\sigma\nu}-g_{\mu\sigma}g_{\rho\nu}\big).
\end{equation}
The Weyl tensor is the trace-free part of the Riemann curvature tensor,
where $R_{\mu\nu\rho\sigma}$ is the Riemann tensor, and $R_{\mu\nu}$ indicates the Ricci tensor.
In fact, for vacuum spacetimes such as the Schwarzschild solution, the Einstein equations imply $R_{\mu\nu}=0$, $R=0$, and the Weyl tensor coincides with the Riemann tensor. In the presence of Weyl and EM contributions, the effective Einstein equations become
\begin{equation}
	G_{\mu\nu}=8\pi G \left(T_{\mu\nu}^{\text{Max}}+\alpha T_{\mu\nu}^{\text{Weyl}}\right),
\end{equation}
where
\begin{equation}
	T_{\mu\nu}^{\rm Max}=F_{\mu\lambda}F_{\nu}^{\lambda}-\frac{1}{4}g_{\mu\nu}F_{\alpha\beta}F^{\alpha\beta}
\end{equation}
and
\begin{equation}
	\begin{aligned}
		T_{\mu\nu}^{\rm Weyl}
		=&\frac{-2}{\sqrt{-g}}\frac{\delta}{\delta g^{\mu\nu}}\int d^4 x\sqrt{-g}C_{\alpha\beta\rho\sigma}F^{\alpha\beta}F^{\rho\sigma}
		\\=&-\frac{1}{2}g_{\mu\nu}C_{\alpha\beta\delta\gamma}F^{\alpha\beta}F^{\delta\gamma}+2C_{\mu\alpha\beta\delta}F_{\nu}^{\alpha}F^{\beta\delta}+2C_{\nu\alpha\beta\delta}F_{\mu}^{\alpha}F^{\beta\delta}\\&-\nabla_{\alpha}\nabla_{\beta}\big(F_{\mu}^{\alpha}F_{\nu}^{\beta}+F_{\nu}^{\alpha}F_{\mu}^{\beta}\big)+g_{\mu\nu}\nabla_{\alpha}\nabla_{\beta}\big(F_{\lambda}^{\alpha}F^{\beta\lambda}\big).
	\end{aligned}
\end{equation}
The nonzero components of the Weyl tensor are 
\begin{equation}\label{AR}
	\begin{array}{cc}
		C_{\hat{t}\hat{r}\hat{t}\hat{r}}=-2\mathcal{A}(r)&
		C_{\hat{t}\hat{\theta} \hat{t}\hat{\theta}}=+\mathcal{A}(r)\\
		C_{\hat{r}\hat{\theta} \hat{r}\hat{\theta}}=-\mathcal{A}(r)&
		C_{\hat{\theta}\hat{\phi}\hat{\theta}\hat{\phi}}=2\mathcal{A}(r)
	\end{array}
	\qquad\qquad\text{where}\qquad \mathcal{A}(r)=\frac{M}{r^3}-\frac{q^2}{r^4}.
\end{equation}
The metric ansatz assumed is
\begin{equation}\label{ds2}
	ds^2=-f(r)dt^2+\frac{dr^2}{f(r)}+R(r)d\Omega^2,
\end{equation}
and the gauge field ansatz, which is static and purely electric, is 
\begin{equation}
	A_{\mu}=(\phi(r),\,0,\,0,\,0),\qquad \text{and} \qquad F_{tr}=-\phi^{'}(r).
\end{equation}
Because the equations are fourth-order and complicated, it is not possible to solve them exactly. Instead we expand $f(r)$, $R(r)$, and $\phi(r)$ to first order in the Weyl correction parameter $\alpha$
\begin{equation}\label{expand}
	\begin{aligned}
		f(r)=&f_{0}(r)+\alpha f_1(r)+\mathcal{O}(\alpha^2),\\
		R(r)=&R_{0}(r)+\alpha R_1(r)+\mathcal{O}(\alpha^2),\\
		\phi(r)=&\phi_{0}(r)+\alpha \phi_1(r)+\mathcal{O}(\alpha^2).
	\end{aligned}
\end{equation}
At zeroth order $(\alpha=0)$, the Maxwell equation gives
\begin{equation}\label{phiprime}
	\phi^{'}_0(r)=\frac{-Q}{r^2}=\frac{-q}{\sqrt{4\pi G}r^2},
\end{equation}
and Einstein equations give $f_0(r)=1-2M/r+q^2/r^2$. This is ordinary Reissner-Nordstr\"{o}m (RN) solution. The final redshift function is \cite{chen2014rotating}
\begin{equation}\label{fr}
	\begin{aligned}
		f(r)&=1-\frac{2 M}{r}+\frac{q^2}{r^2}-\frac{4 \alpha  q^2}{3 r^4}\left(1-\frac{10 M}{3 r}+\frac{26 q^2}{15 r^2}\right),\\
		R(r)&=r^2+\frac{4 \alpha  q^2}{9 r^2}.
	\end{aligned}
\end{equation}
in which $M$ indicates the black hole mass, $q$ refers to the electric charge, and $\alpha$ represents the Weyl correction parameter, which is the central constant in the generalized Einstein-Maxwell action. This parameter dictates the strength of the non-minimal interaction between the Weyl tensor and the electromagnetic field. From a theoretical standpoint, such a coupling is not an arbitrary modification but is motivated by effective field theories, where $\alpha$ represents the leading-order correction arising from higher-curvature corrections that may emerge from vacuum polarization effects in quantum electrodynamics in curved spacetime. Consequently, the presence of $\alpha$ signifies that the propagation and structure of the electromagnetic field are intrinsically linked to the local curvature of spacetime, a departure from the minimal coupling seen in classical RN geometry, and as $\alpha$ tends to zero, the above lapse function approaches the RN black hole lapse function. According to Eq. \eqref{fr}, as $r\to\infty$, the lapse function approaches unity, so $f(r)$ is positive at spatial infinity, and as $r\to 0$, when $q\neq 0$, the dominant term in the lapse function is $(-104\alpha q^4)/(45 r^6)$.
\\The event horizon is determined by the largest positive root of $f(r_+)=0$.
Since the metric functions are obtained perturbatively up to first
order in the Weyl-correction parameter, the near-origin
region should not be used to draw definitive conclusions about the
global number of horizons. In particular, terms proportional to
$\alpha/r^6$ may dominate as $r\to 0$, signaling the breakdown of
the first-order expansion in that region. Therefore, throughout this
work, we identify the physical event horizon as the largest positive
root of the lapse function, restricted to the perturbative domain where
$|\alpha \mathcal{A}(r)|\ll 1$, outside the horizon. For \(\alpha=0\), the standard RN horizon
structure is recovered. For small nonzero \(\alpha\), the Weyl
correction shifts the location of the horizons and may modify the
allowed parameter region. These changes are illustrated in Figure \ref{fig:fr}. In all numerical calculations, we set $M=1$; hence $r$, $q$, and $\alpha$ are presented in units of $M$, $M$, and $M^2$, respectively. Within the perturbative domain, the sign of \(\alpha\) affects the
horizon structure. Negative and positive Weyl corrections shift the
roots of \(f(r)\) in different directions, thereby modifying the
range of charge-to-mass ratios for which a regular event horizon
exists. However, because the solution is perturbative, the number and
location of horizons should be interpreted only within the region
where \(|\alpha \mathcal{A}(r)|\ll1\).
For the spacetime of Eq. \eqref{ds2}, we consider the largest root as the event horizon radius $r_+$. The variation of the horizon radius is displayed in Figure \ref{fig:Horizon}.
Our numerical analysis of the horizon structure reveals that both the electric charge $q/M$ and the Weyl coupling parameter $\alpha/M^2$ contribute to the contraction of the event horizon. As expected, the electric charge $q/M$ exerts the dominant influence, where an increase in its value leads to a significant decrease in the event horizon radius $r_+$. Similarly, we observe that an increase in the Weyl coupling parameter $\alpha/M^2$ also results in a reduction of the horizon size. However, the impact of $\alpha/M^2$ is minimal compared to the primary role played by the electric charge. This suggests that while both parameters effectively reinforce the repulsive potential in the strong-field regime, the numerical shift introduced by the Weyl coupling is relatively small in determining the overall horizon scale.

Having obtained the entropy from the Wald--Iyer Noether-charge prescription \cite{wald1993black}, we now revisit the thermodynamic quantities derived from it. Since the Lagrangian contains the non-minimal curvature coupling $(\alpha C_{\mu\nu\rho\sigma}F^{\mu\nu}F^{\rho\sigma})$, the entropy is not given only by the Bekenstein--Hawking area law.
Therefore, thermodynamic quantities must be computed consistently using the Wald entropy.
\\We work in the fixed charge regime and keep terms up to first order in the Weyl coupling parameter. The mass function obtained from the horizon condition $(f(r_+)=0)$ may be expanded as
\begin{equation}\label{MassW}
M_{\rm W}(r_+)=M_{\rm RN}(r_+)+\alpha M_1(r_+)+\mathcal{O}(\alpha^2),
\end{equation}
where
\begin{equation}
M_{\rm RN}(r_+)=\frac{r_+^2+q^2}{2r_+}, \qquad M_1(r_+)=\frac{2q^2\left(10r_+^2-q^2\right)}{45r_+^5}.
\end{equation}
The first term is precisely the RN mass written in terms of the horizon radius, while the second term is the leading Weyl correction.

\begin{figure}[ht]
\centerline{
		\includegraphics[width=6cm]{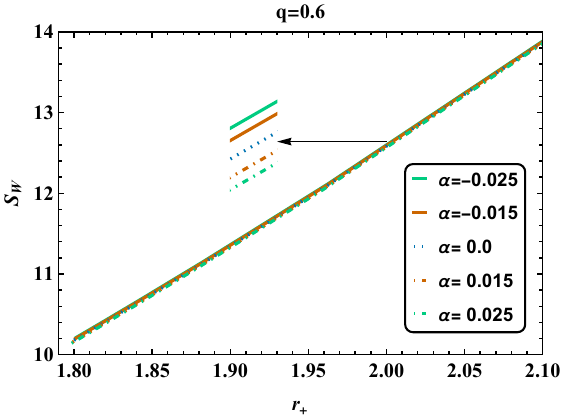} \hspace{-0.2cm}
			\includegraphics[width=6cm]{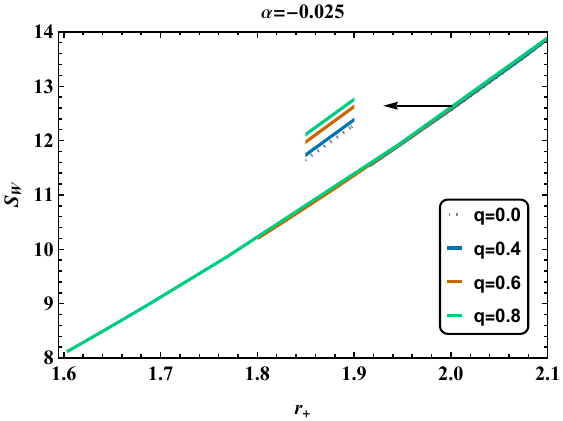} \hspace{0.2cm}}
	\caption{The dependence of the Wald entropy on the charge and the Weyl correction parameter.
		 Left panel: variation with respect to $\alpha$ for fixed $q=0.6$. Right panel: variation with respect to $q$ for fixed $\alpha=-0.025$.
		 }\label{fig:Entropy}
\end{figure}

	In Einstein gravity, where the Lagrangian is linear in the Ricci scalar, the Wald--Iyer Noether-charge entropy reduces to the familiar area law $S_{+}=A_+/4$. In the present case, the Lagrangian contains the non-minimal Weyl--Maxwell coupling
	\begin{equation}\label{entadd}
		\alpha C_{\mu\nu\rho\sigma}F^{\mu\nu}F^{\rho\sigma},
	\end{equation}
	which depends explicitly on the curvature tensor. Therefore, the black hole entropy receives an additional Noether-charge contribution and must be computed using the Wald--Iyer formula rather than the area law alone.
	\\Accordingly,
	the entropy is given by
	\begin{equation}
		S_{\rm W}=
		-2\pi
		\int_{\mathcal H} d^2y\sqrt{h}
		\frac{\partial \mathcal L}
		{\partial R_{\mu\nu\rho\sigma}}
		\epsilon_{\mu\nu}\epsilon_{\rho\sigma},
	\end{equation}
	where $h$ is the determinant of the induced metric on the two-dimensional horizon cross-section $\mathcal{H}$, and $\epsilon_{\mu\nu}$ is the binormal to the horizon normalized as
	$\epsilon_{\mu\nu}\epsilon^{\mu\nu}=-2$.
	The Einstein--Hilbert part of the Lagrangian gives the usual area contribution, while the Weyl--Maxwell coupling produces the first-order correction proportional to $\alpha q^2$ \cite{wald1993black,aros2013wald,iyer1994some,iyer1995comparison}
	\begin{equation}\label{EntW}
		S_{\rm W}=S_{+} + S_{\rm Weyl-EM}.
	\end{equation}
	The horizon area is obtained from $A_+=4\pi R(r_+)$. Thus, the area contribution to the entropy is \cite{bekenstein1973black,bekenstein1994we,jacobson1994black}
	\begin{equation}
		S_{+}
		=\pi\left(r_+^2+\frac{4 \alpha  q^2}{9 r_+^2}\right).
	\end{equation}
	The additional Wald correction comes from Eq. \eqref{entadd} and, to first order in $\alpha$ and using Eq. \eqref{phiprime}, one obtains
	\begin{equation}
		\begin{aligned}
				\left.
				\frac{\partial}{\partial R_{\mu\nu\rho\sigma}}
				\left(\alpha C_{\alpha\beta\gamma\delta}
				F^{\alpha\beta}F^{\gamma\delta}\right)
				\varepsilon_{\mu\nu}\varepsilon_{\rho\sigma}
				\right|_{\mathcal H}&
				=\frac{4\alpha Q^{2}}{3r_{+}^{4}}
				+\mathcal O(\alpha^{2})
			\\
			\Longrightarrow
		S_{\rm Weyl-EM}=-\frac{8\pi\alpha q^{2}}{3 r_{+}^{2}}&
		+\mathcal O(\alpha^{2}).
	\end{aligned}
	\end{equation}
	Therefore, Eq. \eqref{EntW} reduces to
	\begin{equation}\label{SW}
		S_{W}=\pi r_+^2-\frac{20\pi\alpha q^{2}}{9r_{+}^{2}}+\mathcal{O}(\alpha^2).
	\end{equation}
In the limit $\alpha\rightarrow0$, the entropy reduces to the
	standard RN area entropy. The Schwarzschild
	entropy is recovered in the additional neutral limit
	$q\rightarrow0$. Entropy as a function of the parameters $q$ and $\alpha$ is demonstrated in Figure \ref{fig:Entropy}.
	As shown, for a fixed charge parameter, the increase in the Weyl correction parameter from $-0.025$ to $0.025$ induces a monotonic decrease in the black hole entropy. Furthermore, the response of the entropy to the charge parameter exhibits a distinct dependence on the sign of $\alpha$: for negative $\alpha$, increasing $q$ serves to enhance the entropy, whereas for positive $\alpha$, the trend is inverted, with an increase in $q$ leading to a suppression of the entropy value.

\begin{figure}[ht!]
	\centerline{
		\includegraphics[width=6.cm]{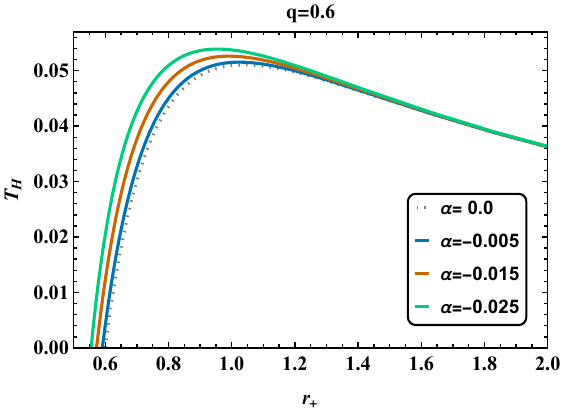} \hspace{0.2cm}
		\includegraphics[width=6.cm]{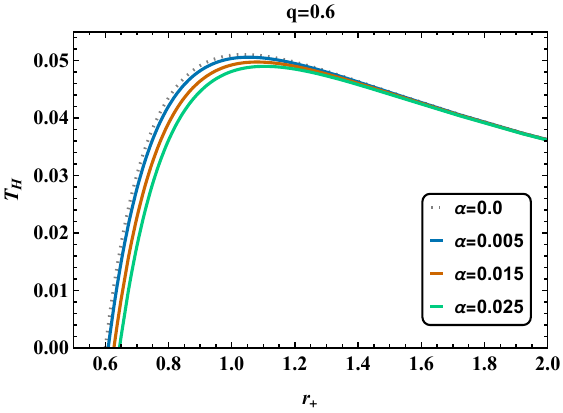} \hspace{-0.2cm}}
			\centerline{
		\includegraphics[width=6.cm]{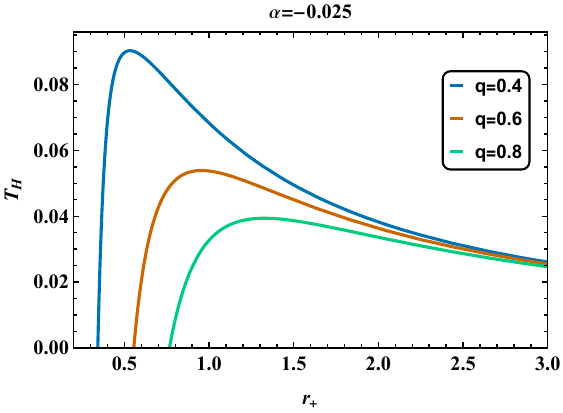} \hspace{0.2cm}
		\includegraphics[width=6.cm]{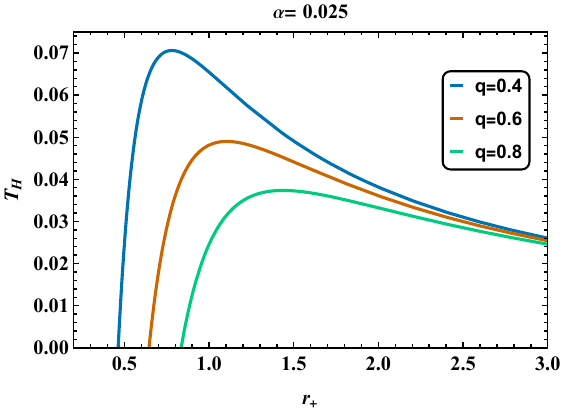} \hspace{-0.2cm}}
	\caption{The Hawking temperature as a function of $r_+$ for different values of $q$ and $\alpha$.}\label{fig:Temp}
\end{figure}

The physical temperature is the Hawking temperature obtained from the surface
gravity,
\begin{equation}
		T_{\rm H}
		=\frac{f'(r_{+})}{4\pi}
		=\frac{r_{+}^{2}-q^{2}}{4\pi r_{+}^{3}}
		+\alpha\left(
		\frac{2q^{4}}{3\pi r_{+}^{7}}
		-\frac{11q^{2}}{9\pi r_{+}^{5}}
		\right)
		+\mathcal O(\alpha^{2}).
	\label{eq:correctedTemperature}
\end{equation}
The Wald formalism modifies the black hole entropy but does not
introduce a second physical temperature distinct from the Hawking
temperature obtained from the surface gravity. As an independent
consistency check, Eqs. \eqref{MassW} and \eqref{SW} give \cite{hawking1974black,hawking1976black}
\begin{equation}
	\left(\frac{\partial M_{\rm W}}{\partial S_{\rm W}}\right)_{q}
	=
	\frac{dM_{\rm W}/dr_{+}}{dS_{\rm W}/dr_{+}}
	=
	T_{\rm H}
	+\mathcal{O}(\alpha^{2}).
	\label{eq:first-law-check}
\end{equation}
It is evident that in the limit as $\alpha$ approaches zero, the above relation converges to $(r_+^2-q^2)/(4 \pi  r_+^3)$, which corresponds to the Hawking temperature of the RN black hole. Similarly, in the limit as $q$ approaches zero, it converges to the Hawking temperature of the Schwarzschild black hole. In Figure \ref{fig:Temp}, the Hawking temperature curve is plotted against the horizon radius for various values of $\alpha$ and $q$.
The Hawking temperature profile reveals a thermodynamic phase transition, characterized by a maximum point in the $T_+-r_+$ diagram. Our results indicate that both the electric charge and the Weyl parameter act as controlling factors for this critical behavior. Specifically, as we increase $q$ or shift $\alpha$ towards positive values, the critical point shifts to a larger horizon radius but drops to a lower temperature value.

\begin{figure}[ht!]
	\centerline{
		\includegraphics[width=7cm]{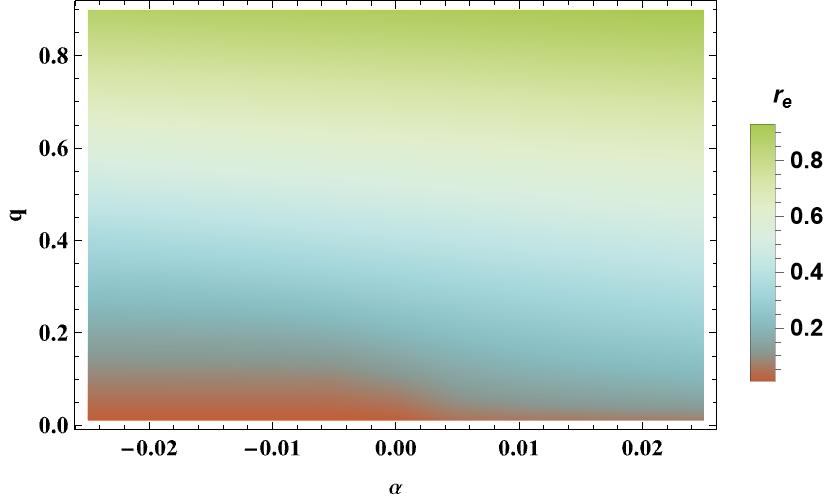} \hspace{0.5cm}}
	\caption{The extremal radius as a function of the parameters $\alpha$ and $q$.}\label{fig:Rem}
\end{figure}

The extremal radius $r_e$ is defined by the zero-temperature condition $T_{\rm H}(r_e)=0$. If the evaporation process approaches this zero-temperature configuration, it may be interpreted as a candidate remnant  \cite{hamil2022effect,karmakar2022quasinormal,hosseinifar2026study}. For fixed $q>0$ and $|\alpha|/q^{2}\ll1$, the RN-connected solution is $ r_e = q+10\alpha/(9q) +\mathcal{O}\!\left(\alpha^{2}/q^{3}\right). $ This expansion is nonuniform in the neutral limit and therefore cannot be extrapolated to $q=0$.
Figure \ref{fig:Rem} illustrates the behavior of the extremal radius with respect to the parameters $\alpha$ and $q$.
Our numerical analysis demonstrates that for a fixed Weyl coupling $\alpha$, the extremal radius increases monotonically with the electric charge. This indicates that a higher charge strengthens the repulsive mechanism, halting the collapse at a larger physical scale. Furthermore, we observe that the extremal radius is a strictly increasing function of the Weyl parameter $\alpha$ within the investigated range. Specifically, as $\alpha$ is swept from $\alpha = -0.025$ to $\alpha = 0.025$, $r_{\rm e}$ undergoes a continuous increase. This suggests that a positive Weyl coupling reinforces the effective repulsion in the strong-field regime, leading to a larger extremal radius. In contrast, a negative coupling leads to a more compact final state. It is worth noting that while $\alpha$ introduces a clear modification to the $r_{e}$ structure, its influence remains secondary to the dominant role played by the electric charge.

	Since the metric functions have been obtained only to first order in the
	Weyl-coupling parameter $\alpha$, the numerical analysis must be restricted
	to a domain in which the Weyl correction remains perturbatively small
	throughout the exterior spacetime. The relevant curvature function at
	zeroth order is introduced in Eq. \eqref{AR}.
	The same function appears in the polarization-dependent factors of the
	effective photon metric later in Section \ref{Sec2} Eq. \eqref{WPM}.
	Therefore, the factors $1\pm16\alpha \mathcal{A}(r)$ must remain sufficiently
	far from zero at every point outside the event horizon. We consequently
	define the dimensionless control parameter
	\begin{equation}
			\epsilon_{\rm W}
			=
			16|\alpha|
			\max_{r\geq r_{+}^{(0)}}
			\left|\mathcal{A}(r)\right|
			<\delta,
			\qquad
			\delta=0.1 ,
		\label{eq:Weyl-validity}
	\end{equation}
	where $
		r_{+}^{(0)}
		=
		M+\sqrt{M^{2}-q^{2}}$
	is the outer RN horizon. Thus, the maximization in
	Eq.~\eqref{eq:Weyl-validity} is performed over the complete exterior
	radial interval $
		r_{+}^{(0)}\leq r<\infty$.
	Using the zeroth-order horizon in this condition is consistent with our
	first-order treatment. Indeed, the $\alpha$-dependent displacement of
	the horizon produces only an $\mathcal{O}(\alpha^{2})$ contribution to
	the quantity $\alpha A\!\left(r_{+}^{(0)}\right)$.
	
	To evaluate the maximum analytically, we introduce the dimensionless
	variables
	\begin{equation}
		x=\frac{r}{M},
		\qquad
		q_m=\frac{q}{M},
		\qquad
		\lambda=\frac{\alpha}{M^{2}} .
	\end{equation}
	The dimensionless curvature function and the outer horizon are then
	\begin{equation}
		M^{2}\mathcal{A}(r)
		=
		a(x,q_m)
		=
		\frac{x-q_m^{2}}{x^{4}},
		\qquad
		x_{+}=1+\sqrt{1-q_m^{2}}, \qquad x\geq x_{+}.
	\end{equation}
	Its radial derivative is
	\begin{equation}
		\frac{\partial a}{\partial x}
		=
		\frac{4q_m^{2}-3x}{x^{5}} .
	\end{equation}
	The stationary point is located at $x_{*}=4q_m^{2}/3$. For the parameter
	range used in our numerical calculations, $0\leq q_m\leq0.9$, this point
	lies inside the outer horizon. Consequently, $\mathcal{A}(r)$ decreases
	monotonically in the exterior region, and its maximum is attained at
	the outer horizon:
	\begin{equation}
		\max_{r\geq r_{+}^{(0)}}M^{2}|\mathcal{A}(r)|
		=
		\frac{\sqrt{1-q_m^{2}}}
		{\left(1+\sqrt{1-q_m^{2}}\right)^{3}} .
	\end{equation}
	The largest value of this expression over $0\leq q_m\leq0.9$ is
	\begin{equation}
		\max_{\substack{r\geq r_{+}^{(0)}\\0\leq q/M\leq0.9}}
		M^{2}|\mathcal{A}(r)|
		=
		\frac{4}{27},
	\end{equation}
	which occurs at
	\begin{equation}
		\frac{q}{M}=\frac{\sqrt{3}}{2},
		\qquad
		r=r_{+}^{(0)}=\frac{3M}{2}.
	\end{equation}
	
\begin{figure}[ht]
	\centerline{
		\includegraphics[width=7cm]{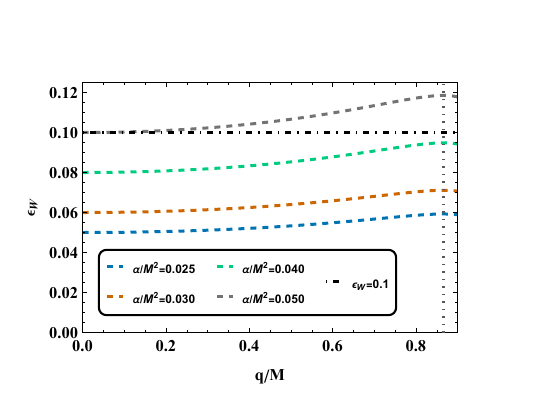} \hspace{-0.2cm}}
	\caption{Maximum perturbative-control parameter $\epsilon_{\rm W} =16|\alpha|\max_{r\geq r_{+}^{(0)}}|\mathcal{A}(r)|$ as a function of $q/M$. The curves with $|\alpha|/M^{2}=0.030$, $0.040$, and $0.050$ are shown only to demonstrate the approach to, or violation of, the operational bound $\epsilon_{\rm W}=0.1$ and are not used in the physical analysis. All subsequent numerical results are restricted to $|\alpha|/M^{2}\leq0.025$ and $0\leq q/M\leq0.9$.} \label{fig:weyl-validity}
	\label{fig:pcp}
\end{figure}

	Although the condition $\epsilon_{\rm W}<0.1$ by itself gives
	$|\alpha|/M^{2}<27/640\simeq0.0422$, we adopt the more conservative
	numerical domain
	\begin{equation}
			-0.025\leq\frac{\alpha}{M^{2}}\leq0.025,
			\qquad
			0\leq\frac{q}{M}\leq0.9 .
	\end{equation}
	Within this domain, the largest possible value of the perturbative
	control parameter is
	\begin{equation}
		\epsilon_{\rm W}^{\rm max}
		=
		16(0.025)\frac{4}{27}
		=
		0.05926<0.1.
	\end{equation}
	To ensure the reliability of our perturbative approach and to maintain the validity of the Weyl corrections, we analyze the behavior of the control parameter $\epsilon_{\rm W}$ across the parameter space. Figure \ref{fig:weyl-validity} illustrates $\epsilon_{\rm W}$ as a function of $q/M$ for various values of the Weyl coupling parameter.
The vertical dotted line marks $q/M=\sqrt{3}/2$, at which the exterior maximum of the curvature factor, and consequently $\epsilon_{\rm W}$ for fixed $|\alpha|/M^{2}$, reaches its largest value over the adopted charge interval.
While curves for $|\alpha|/M^{2}=0.030$, $0.040$, and $0.050$ are displayed to demonstrate how the system approaches or eventually violates this perturbative limit, these values are excluded from our physical analysis. Consequently, to stay within a well-behaved regime, all subsequent numerical results and physical discussions in this work are restricted to the conservative range of $|\alpha|/M^{2}\leq 0.025$ and $0\leq q/M\leq 0.9$.
	We emphasize that Eq.~\eqref{eq:Weyl-validity} controls the
	perturbative reliability of the Weyl expansion. The existence of an
	RN-connected outer horizon and the positivity of the corrected Wald
	entropy and Hawking temperature are additionally verified
	point by point. Parameter values that fail any of these conditions are
	excluded from the physical domain.

\begin{figure}[ht!]
	\centerline{
		\includegraphics[width=6.cm]{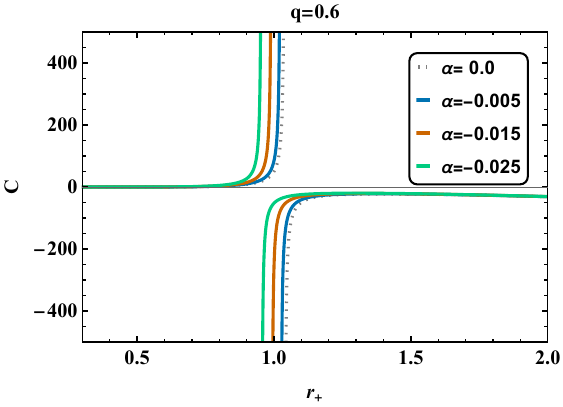} \hspace{0.2cm}
		\includegraphics[width=6.cm]{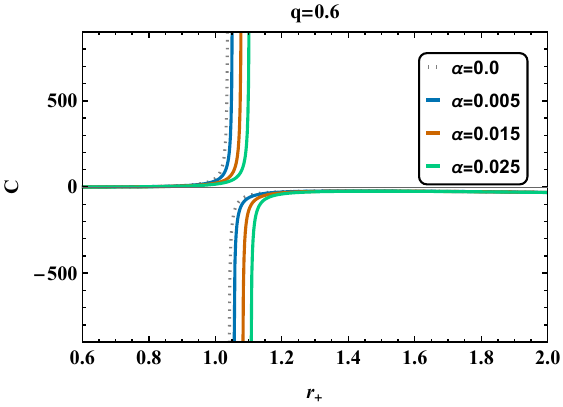} \hspace{-0.2cm}}
			\centerline{
		\includegraphics[width=6.cm]{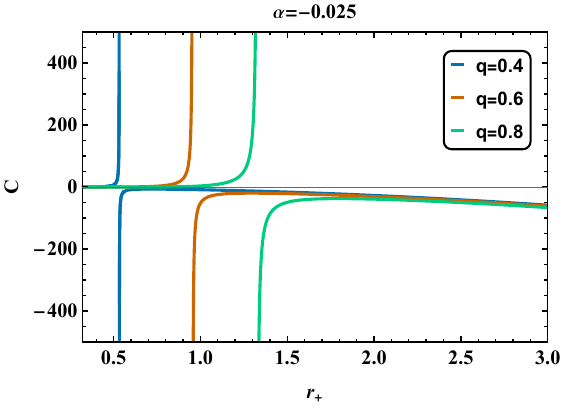} \hspace{0.2cm}
		\includegraphics[width=6.cm]{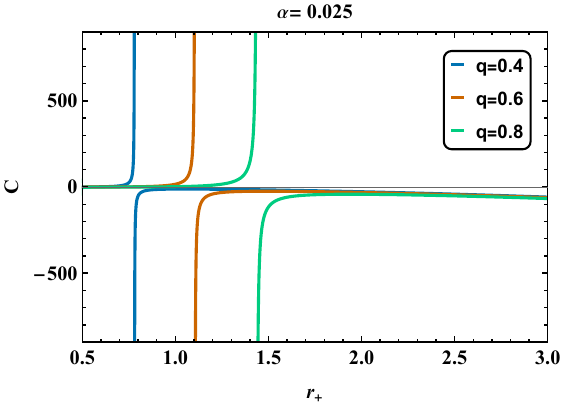} \hspace{-0.2cm}
		}
	\caption{The heat capacity as a function of $r_+$ for various choices of $q$ and the Weyl correction parameters.}\label{fig:Cv}
\end{figure}

The heat capacity of the black hole at fixed charge is given by \cite{brown1994temperature,brill1997thermodynamics}
\begin{equation}
	\begin{aligned}
	C&=\left(\frac{\partial M_W}{\partial T_{\rm H}}\right)_q\\
	&=\frac{2 \pi  r^2 \left(4 \alpha  q^4-3 q^2 \left(3 r^4+8 \alpha  r^2\right)+9 r^6\right)}{-168 \alpha  q^4+q^2 \left(27 r^4+220 \alpha  r^2\right)-9 r^6}+\mathcal{O}(\alpha^2).
	\end{aligned}
\end{equation}
This is a better first-order expression for discussing phase transition. The phase transition point is determined by
\begin{equation}
	T'_{\rm H}\big|_{r=r_c}=0
\end{equation}
which yields
\begin{equation}
	9 r_c^6 -q^2 \left(27 r_c^4+220 \alpha  r_c^2\right) + 168 \alpha  q^4=0.
\end{equation}
For small $\alpha$, the critical radius is shifted by the correction from the RN value $r_c^{\rm RN}=\sqrt{3}q$ to
\begin{equation}
	r_c=\sqrt{3}q+\alpha\frac{82 }{27 \sqrt{3} q}+\mathcal{O}(\alpha^2).
\end{equation}
Although the heat capacity can formally be expanded as 
\begin{equation}
	C=
	C_{\rm RN}
	+
	\alpha C_1
	+
	\mathcal{O}(\alpha^2).
\end{equation}
This expansion is not uniform near the critical point where  $T'_{\rm RN}(r_+)=0$. Therefore, to study the divergence and the change of sign of the heat capacity, we keep the first-order numerator and denominator unexpanded.
Figure \ref{fig:Cv} presents the heat capacity for different values of parameters $\alpha$ and $q$.
Consistent with the behavior observed in the Hawking temperature profile, the heat capacity exhibits a discontinuity, signalling a phase transition. For smaller horizon radii, the heat capacity is positive; as the horizon radius increases, the black hole's heat capacity diverges at a critical point and then changes to negative. Since a positive heat capacity indicates that the black hole's temperature rises with the absorption of heat, while a negative heat capacity implies that the black hole cools despite absorbing heat, it can be concluded that the black hole is thermodynamically stable for smaller horizon radii. As the horizon radius increases, a phase transition occurs, leading to a thermodynamically unstable state \cite{dong2025some}.
This divergence point marks the boundary between the locally stable and unstable phases of the black hole. Our results show that the location of this phase transition is sensitive to the Weyl correction parameter. Specifically, for a fixed charge $q$, shifting $\alpha$ from $-0.025$ to $0.025$ causes the divergence point to migrate toward a larger horizon radius. This shift indicates that positive Weyl corrections expand the regime of thermodynamic instability, requiring a larger geometric scale for the black hole to achieve a transition between its physical states.

The final thermodynamic quantity studied in this section is the free energy of the black hole, which is calculated using Eqs. \eqref{MassW}, \eqref{SW}, and \eqref{eq:correctedTemperature} as follows \cite{chamblin1999holography}
\begin{equation}
	\begin{aligned}
		F&=M_{\rm W}-T_{\rm H} S_{\rm W}.
	\end{aligned}
\end{equation}
Using the Wald entropy and the Hawking temperature, the first-order expansion gives
\begin{equation}
F=
F_{\rm RN}
+
\alpha F_1
+
\mathcal{O}(\alpha^2),
\end{equation}
where
\begin{equation}
F_{\rm RN}=
\frac{r_+^2+3q^2}{4r_+},\qquad
F_1=
\frac{  100 q^2 r_+^2-57 q^4}{45 r_+^5}.
\end{equation}
For $\alpha\rightarrow0$, this expression reduces to the RN black hole free energy, while in the neutral limit $q\rightarrow0$, it gives the Schwarzschild result $F_{\rm Sch}=r_+/4$. Therefore, the complete first-order thermodynamic structure is obtained from the Wald entropy, which explicitly separates the ordinary RN contribution from the leading Weyl correction.
\section{Topological Charge of Thermodynamic Potentials}\label{Sec6}
Investigating black holes from a topological perspective offers key insights into their thermodynamic behavior. Although Section \ref{Sec4} focused on the thermodynamic phase transitions, the topological classification of the critical points remained unexplored. In this section, we utilize a topological scheme to determine the nature of these critical points. Specifically, we identify the conventional and novel types for the temperature critical points \cite{alipour2023topological,wei2022topology}, and further classify the topological structure of the free energy landscape.
\\To explore the critical point of the Hawking temperature, we define a temperature-dependent potential in the form \cite{wei2022topology}
\begin{eqnarray}
	\Phi =\frac{1}{\sin \theta} T_{\rm H}.
\end{eqnarray}
The vector space of this potential can be represented using the $\phi^{\Phi}_r=\partial_{r_{+}}\Phi,\,
\phi^{\Phi}_\theta=\partial_\theta\Phi$ vectors.
the above vectors are being normalized from $n^\Phi_i=\phi^\Phi_{i}/||\phi||$. 
The points at coordinates $(\phi_r^{\Phi}=0, \pi/2)$ represent the zeros of the vector space. Using a topological approach, these points can be viewed as defects in the vector space, to which a topological number can be assigned based on the rotation of vectors around these points.
To achieve this, a closed contour is drawn around each arbitrary point $r_i$, and by employing a change of variables $r=a \cos\vartheta +r_i$ and $\theta=b \sin\vartheta+\frac{\pi}{2}$, the following integral is calculated to determine the topological charge enclosed within the closed loop $c_i$
\begin{eqnarray}\label{winding}
	w_i=\frac{1}{2\pi}\oint_{c_i} d(\arctan\frac{n_\theta}{n_r}).
\end{eqnarray}
If the contour does not enclose any zero of the vector field, the
	winding number integral vanishes. A contour enclosing one isolated
	nondegenerate zero gives $w_i=+1$ when the vector rotates once
	counterclockwise and $w_i=-1$ when it rotates once clockwise. If
	several zero points are enclosed, the winding number equals the sum
	of their individual topological charges.
To evaluate the thermodynamic topology associated with the Hawking temperature, we calculated the topological charge using the integral from Eq. \eqref{winding} over closed contours $c_i$.

\begin{figure}[ht!]
    \centerline{
		\includegraphics[width=6cm]{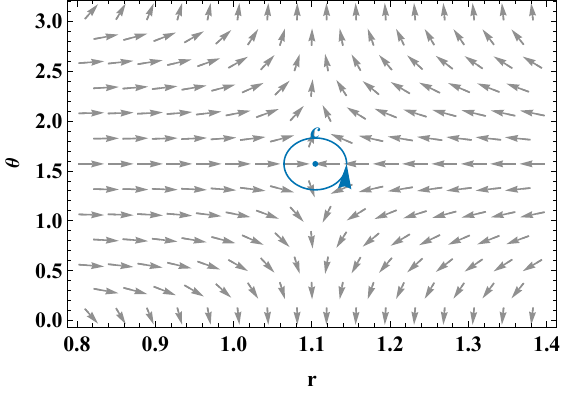} \hspace{0.5cm}
		\includegraphics[width=6cm]{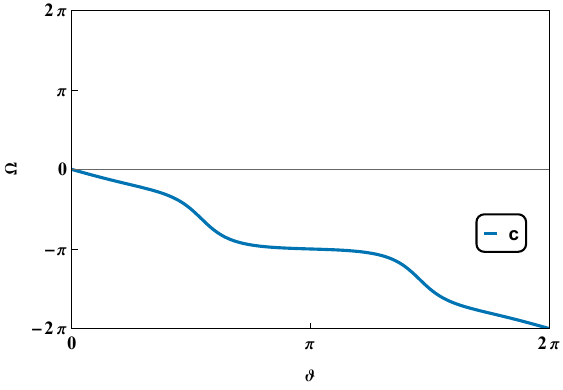}}
	\caption{Normalized vector field $n^\Phi$ and topological defect for the Hawking temperature, choosing $q=0.6$, $\alpha=0.025$, and contour semi-axes $a=0.3$, and $b=0.3$.}\label{fig:TopoTH}
\end{figure}

For $q=0.6$ and $\alpha=0.025$, the resulting vector field is illustrated in Figure \ref{fig:TopoTH}, along with the corresponding topological charge of its zero point.
Considering that the topological charge at the zero point is $-1$, it indicates that this black hole has only one critical point of the conventional type at Hawking temperature \cite{wei2022topology}. 
\\Our topological analysis of the Hawking temperature for different values of $\alpha$ and $q$ reveals a robust topological behavior. Specifically, although the location of the zero point shifts with variations in these parameters, the topological charge and the number of critical points remain invariant. This result is consistent with our analysis of the black hole temperature presented in the previous section.

\begin{figure}[ht!]
		\centerline{
		\includegraphics[width=6cm]{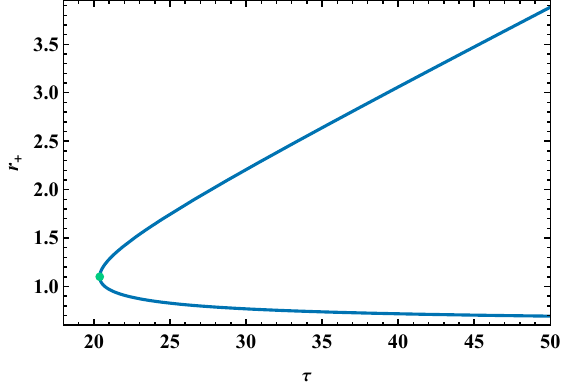}}
	\caption{The horizon radius as a function of parameter $\tau$ for $q=0.6$ and $\alpha=0.025$.}\label{fig:TopoTau}
\end{figure}

\begin{figure}[ht!]
		   \centering
		\includegraphics[width=6cm]{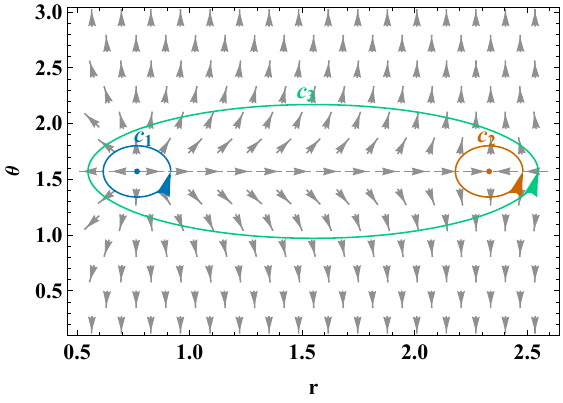} \hspace{0.5cm}
		\includegraphics[width=6cm]{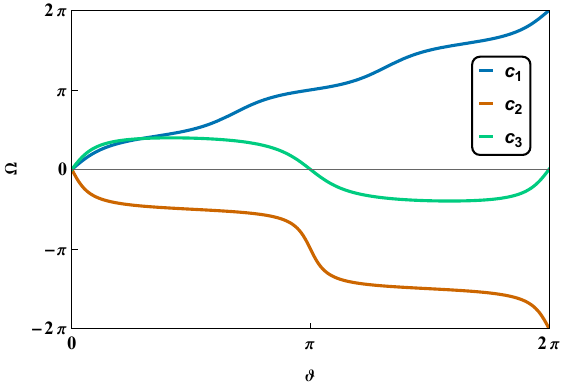} \hspace{-0.2cm}\\
	\caption{Normalized vector field $n^{\mathcal{F}}$ and topological charge of generalized free energy for $q=0.6$, $\alpha=0.025$, and $\tau=10\pi$. Contours $c_1$ and $c_2$ enclose the zero points with topological charges $+1$ and $-1$, respectively, whereas contour $c_3$ encloses an arbitrary regular point yielding a net charge of $w = 0$.}\label{fig:TopoFr}
\end{figure}

The next quantity analyzed from a topological perspective is the generalized free energy outside the cavity, which enables the classification of black hole states. By employing Eqs. \eqref{MassW} and \eqref{SW}, the generalized free energy is defined as follows \cite{wei2022black,afshar2025topological}
\begin{equation}
	\mathcal{F}=M_{\rm W}-\frac{S_{\rm W}}{\tau},
\end{equation}
where $\tau$ represents the inverse of temperature outside the cavity.
Using the vectors $\phi_r^{\mathcal{F}}=\partial_{r_+}\mathcal{F},\, \phi_\theta^{\mathcal{F}}=-\cot \theta \csc \theta$, this potential field can be represented in vector space. The zero points of this vector space are located at $\partial_{r_+}\mathcal{F}=0$. Therefore, we can find a relation for $\tau$ as a function of the horizon radius
which, up to first order numerator–denominator form in $\alpha$, is expressed as
\begin{equation}
	\tau=
	\frac{4 \pi  \left(20 \alpha  q^2 r^3+9 r^7\right)}{4 \alpha  q^4-3 q^2 \left(3 r^4+8 \alpha  r^2\right)+9 r^6
	}+\mathcal{O}(\alpha^2).
\end{equation}
The first term denotes the inverse of the RN black hole temperature.
In Figure \ref{fig:TopoTau}, the variations of the curve $r_{+}$ against $\tau$ are illustrated.
For the choice of $\alpha=0.025$ and $q=0.6$, it is evident that the curve has two branches for $\tau>\tau_c=20.38$.
\\In Figure \ref{fig:TopoFr}, with the selection of $\alpha=0.025,\,q=0.6$ and $\tau=10\pi$, the vector space of the potential $\mathcal{F}$ is depicted.
It is clear that in this vector space, there are two zero points enclosed by contours $c_1$ and $c_2$, with the corresponding topological charges of $+1$ and $-1$, respectively. 

Now, we focus on the universal topological classification, which includes four classes as $W^{1-}$: unstable small and unstable large, $W^{0+}$: stable small and unstable large, $W^{0-}$: unstable small and stable large, and $W^{1+}$: stable small and stable large \cite{wei2024universal,wu2025novel,rizwan2025universal}. To accomplish this, we calculate the total topological charge from $W=\sum_{i} w_i$ and examine the behavior of $\beta=1/T_+$ at the minimum horizon $r_m$ and at infinity.

Given that $\beta(r_m)=\infty$ and $\beta(\infty)=\infty$ are satisfied, and $w_1=1$ and $w_2=-1$, this indicates that the black hole belongs to class $W^{0+}$.

Furthermore, we observe that for both the classical limit ($\alpha\to 0$) and the negative Weyl coupling regime, the results remain qualitatively consistent with those of the positive regime. Specifically, the number of zero points (and their corresponding topological charges) remains invariant under the variation of the Weyl parameter. This constancy implies that the global phase structure and the arrangement of topological defects are robust against Weyl-type corrections, further confirming that the black hole consistently belongs to the $W^{0+}$ topological class across the investigated parameter space.
\section{The Effective Metric and Shadow}\label{Sec2}
The trajectory of massless particles around the black hole is a crucial area of research in astrophysics.
A black hole shadow is essentially formed when light from distant sources is obstructed by the gravitational pull of the black hole, creating a visually dark disk \cite{falcke2000shadow,perlick2015influence}.

For Weyl-Maxwell coupling, the modification shows up as a polarization-dependent rescaling of the angular part.\\
A convenient and widely used form is
\begin{equation}\label{dseff}
	ds^2_{ph,\pm}=-f(r)dt^2+\frac{dr^2}{f(r)}+\frac{R(r)}{W_{\pm}(r)}d\Omega^2.
\end{equation}
Thus, the primary modification with respect to the standard
metric is that the area radius squared seen by photons becomes
\begin{equation}\label{Reff}
	R_{\rm eff \pm}(r)=\frac{R(r)}{W_{\pm}(r)}.
\end{equation}
For a spherically symmetric background, $W_{\pm}(r)$ is controlled by the (background) Weyl tidal amplitude. A practical way to write it is
\begin{equation}\label{WPM}
	W_{\pm}(r)=\frac{1\mp 8 \alpha \mathcal{A}(r)}{1\pm 16\alpha \mathcal{A}(r)}.
\end{equation}
The perturbative-validity condition introduced in Eq. \eqref{eq:Weyl-validity} guarantees
	that the denominators of both polarization factors remain strictly
	positive throughout the exterior region. Indeed, within the adopted
	parameter domain,
	\begin{equation*}
		\frac{|\alpha|}{M^{2}}\leq0.025,
		\qquad
		0\leq\frac{q}{M}\leq0.9,
	\end{equation*}
	we have
	\begin{equation*}
		16|\alpha|
		\max_{r\geq r_{+}^{(0)}}|\mathcal{A}(r)|
		\leq
		16(0.025)\frac{4}{27}
		=
		0.05926.
	\end{equation*}
	Consequently,
	\begin{equation*}
		1\pm16\alpha \mathcal{A}(r)
		\geq
		1-0.05926
		=
		0.94074>0,
		\qquad
		r\geq r_{+}^{(0)}.
	\end{equation*}
	Thus, neither $W_{+}(r)$ nor $W_{-}(r)$ develops a pole in the
	complete exterior region considered in this work.
For RN-type curvature, the tidal amplitude $\mathcal{A}(r)$ is introduced in Eq. \eqref{AR}.
Defining $s=+1$ for the $W_+$ branch and $s=-1$ for the $W_-$ branch, we may write
\begin{equation}
W_s(r)=\frac{1-8s\alpha \mathcal{A}(r)}
{1+16s\alpha  \mathcal{A}(r)}.
\end{equation}
To first order in \(\alpha\), $W_s(r)=1-24s\alpha  \mathcal{A}(r)+\mathcal O(\alpha^2)$,
and therefore
\begin{equation}
\frac{1}{W_s(r)}
=
1+24s\alpha \mathcal{A}(r)+\mathcal O(\alpha^2),
\end{equation}
and the effective angular radius seen by photons introduced in Eq. \eqref{Reff} can be rewritten as $R_{\rm eff}^s(r)=R(r)/W_s(r)$.
Using Eq. \eqref{fr}, we expand this quantity perturbatively up to first order in the Weyl coupling parameter, yielding
\begin{equation}\label{Reff2}
R_{\rm eff}^{s}(r)=r^2+\alpha \rho^s(r)+\mathcal O(\alpha^2),
\end{equation}
where the linear Weyl correction term $\rho^s(r)$ is given by
\begin{equation}\label{rhos}
\rho^s(r)=\frac{24sM}{r}
+
\left(\frac{4}{9}-24s\right)\frac{q^2}{r^2}.
\end{equation}
Hence, evaluating Eq. \eqref{rhos} for the positive and negative polarization branches yields
\[
\rho^+(r)=\frac{24M}{r}-\frac{212q^2}{9r^2},
\qquad
\rho^-(r)=-\frac{24M}{r}+\frac{220q^2}{9r^2}.
\]
Now, to study the path of light rays in the equatorial plane $(\theta=\pi/2)$, we employ the Lagrangian method as \cite{schneider2018shadow}
\begin{eqnarray}\label{LagSh}
	\mathcal{L}(x,\,\dot{x})=\frac{1}{2}\left(-f(r)\dot{t}^2+\frac{\dot{r}^2}{f(r)}+R_{\rm eff}^s(r)\dot{\phi}^2\right),
\end{eqnarray}
in which coordinates are shown as $x^\mu=(t,\,r,\,\phi)$, $g_{\mu\nu}$ refers to the metric tensor, and $\dot{x}^{\mu}$ represent the derivatives of the coordinates
\begin{eqnarray}\label{dLagsh}
	\frac{d}{d\lambda}\left(\frac{\partial\mathcal{L}}{\partial\dot{x}^\mu}\right)-\frac{\partial\mathcal{L}}{\partial x^\mu}=0.
\end{eqnarray}
Using the Euler-Lagrange equation and defining two constants of motion, $E=-\partial_{\dot{t}}\mathcal{L},\, L =\partial_{\dot{\phi}}\mathcal{L}$, we can derive the equations of motion from
\begin{align}\label{trajj}
	\frac{1}{2}\dot{r}^2+\frac{1}{2}\left(\delta^m+\frac{\left.L\right.^2}{R_{\rm eff}^s(r)}\right)f(r)=\frac{1}{2}E^2.
\end{align}
Here, $\delta^m = 0$ corresponds to null geodesics (massless particles), whereas $\delta^m = 1$ describes timelike geodesics (massive particles). Therefore, for massless particles, we can express it as
\begin{align}\label{trajsh}
	\frac{1}{2}\dot{r}^2+V^l_{\rm eff}(r)=\frac{1}{2}E^2.
\end{align}
Thus, the null effective potential is expressed as
\begin{equation}
V_{\rm eff}^{s}(r)
=
\frac{\left.L\right.^2}{2}
\frac{f(r)}{R_{\rm eff}^{s}(r)}.
\end{equation}
Expanding $f(r)=f_0(r)+\alpha f_1(r)$, one finds
\begin{equation}
V_{\rm eff}^{s}(r)
=
V_{\rm RN}(r)
+
\alpha V_1^{s}(r)
+
\mathcal O(\alpha^2),
\end{equation}
where
\begin{equation}
V_{\rm RN}(r)
=
\frac{\left.L\right.^2}{2r^4}
\left(r^2-2Mr+q^2\right),
\qquad
V_1^{s}(r)
=
\frac{\left.L\right.^2}{2}
\left[
\frac{f_1(r)}{r^2}
-
\frac{f_0(r)\rho^s(r)}{r^4}
\right].
\end{equation}
Therefore, for the two polarizations, this gives
\begin{equation}
	\begin{aligned}
V_1^{+}(r)&
=
\frac{\left.L^+\right.^2}{90r^8}
\left[
2160M^2r^2
-3000Mq^2r
-1080Mr^3
+956q^4
+1000q^2r^2
\right],
\\
V_1^{-}(r)&
=
\frac{\left.L^-\right.^2}{90r^8}
\left[
-2160M^2r^2
+3480Mq^2r
+1080Mr^3
-1204q^4
-1160q^2r^2
\right].
\end{aligned}
\end{equation}
Considering that for a stable circular orbit, the conditions $\dot{r}=0$ and $\ddot{r}=0$ must be satisfied, the photon radius can be calculated from the relation \cite{virbhadra2000schwarzschild,claudel2001geometry}
\begin{equation}\label{photon}
	\bigg(\frac{R_{\rm eff}^s(r)}{f(r)}\bigg)^{'}_{r=r_{\rm ph}^{s}}=0.
\end{equation}
The above equation is the photon-sphere equation.
\\Working to leading order in $\alpha$ for the RN background with Weyl-Maxwell photon metric, we can write the outer-sphere radius as
\begin{equation}\label{rphoton}
	r_{\rm ph}^{s}= r_0+\alpha r_{1}^s+\mathcal{O}(\alpha^2).
\end{equation}
\\For the zero order $\alpha=0$ the outer photon sphere radius solves $\partial_r \left(r^2/f(r)\right)=0$, so the outer solution is 
\begin{equation}
	r_0=\frac{3M+\sqrt{9M^2-8q^2}}{2}.
\end{equation}
(There is also an inner root with the minus sign, typically inside/near the inner horizon and not used for the shadow.)
\\
The first-order correction can be decomposed as
\begin{equation}
r_1^{s}=\Delta r_{\rm bg}+s\Delta r_{\rm pol},
\end{equation}
where
\begin{equation}
\Delta r_{\rm bg}
=
\frac{
	8q^2
	\left(
	52q^2+25r_0^2-90Mr_0
	\right)
}
{
	45r_0^3(r_0^2-2q^2)
},\qquad\Delta r_{\rm pol}
=
\frac{12(Mr_0-q^2)}{r_0^3}.
\end{equation}
Thus, the explicit leading-order radii are expressed as
\begin{equation}
r_{\rm ph}^{+}
=
r_0
+
\alpha
\left[
\Delta r_{\rm bg}
+
\Delta r_{\rm pol}
\right]
+
\mathcal O(\alpha^2),
\qquad
r_{\rm ph}^{-}
=
r_0
+
\alpha
\left[
\Delta r_{\rm bg}
-
\Delta r_{\rm pol}
\right]
+
\mathcal O(\alpha^2).
\end{equation}
When $\alpha\to 0$, both branches' radii tend to the standard RN photon $r_0$.

\begin{figure}[ht!]
		\centerline{
		\includegraphics[width=7cm]{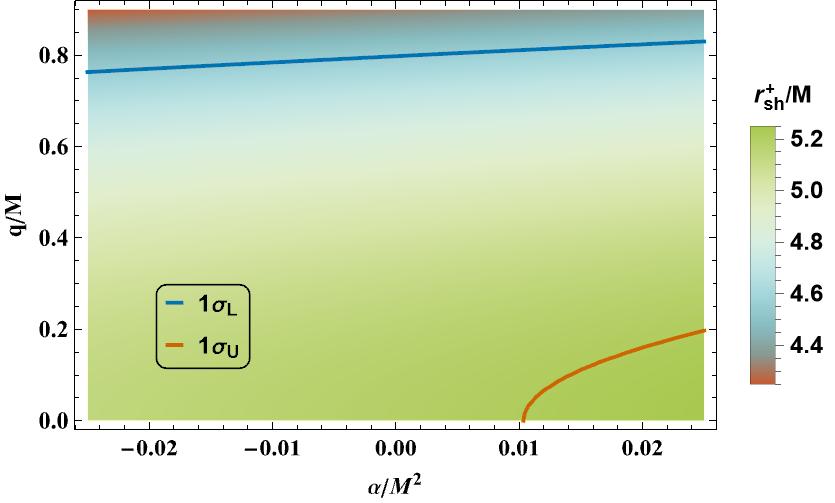} \hspace{0.5cm}
		\includegraphics[width=7cm]{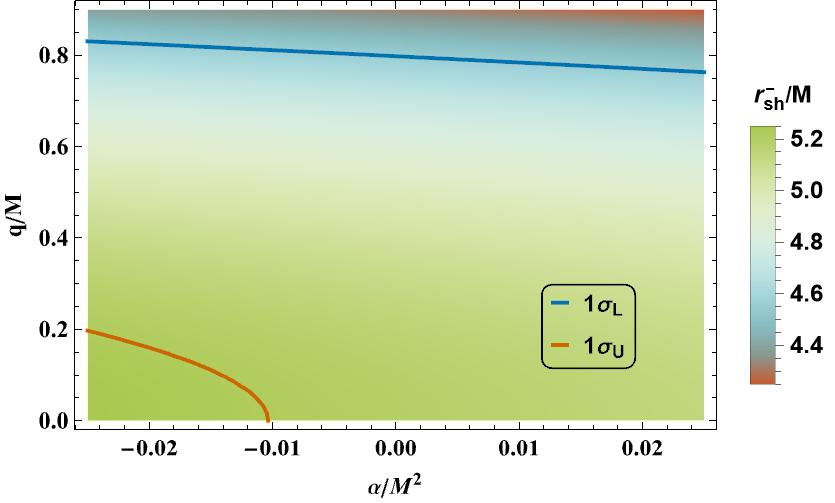}}
	\caption{Black hole shadow radius in the $(\alpha/M^2, q/M)$ parameter space. The colored lines represent the boundaries of $1\sigma$ region from EHT measurements of Sgr A$^*$. The subscripts 'L' and 'U' denote the lower and upper bounds, respectively.}\label{fig:Shadow}
\end{figure}

The shadow radius represents the critical impact parameter of photons as observed by a distant observer.
It is determined by the observer's position and is fundamentally influenced by the gravitational effects of the black hole. Since the lapse function of the black hole is asymptotically flat, the shadow of the black hole can be calculated using \cite{perlick2015influence}
\begin{equation}\label{shadow}
	\begin{aligned}
		r_{\rm sh}^{s}&=\frac{\sqrt{R_{\rm eff}^{s}(r)}}{\sqrt{f(r)}}\bigg|_{r=r_{\rm ph}^{s}}.
	\end{aligned}
\end{equation}
To first order, the black hole shadow is expressed as
\begin{equation}
\left(r_{\rm sh}^{s}\right)^2
=
b_0^2+\alpha B_1^{s}+\mathcal O(\alpha^2),
\end{equation}
where
\begin{equation}
b_0^2=\frac{2r_0^3}{r_0-M},
\qquad
B_1^{s}
=
24s
-
\frac{4(r_0-3M)^2}{5(r_0-M)^2}.
\end{equation}
Therefore, the shadow for the positive and negative polarizations is given by
\begin{equation}
	\begin{aligned}
\left(r_{\rm sh}^{+}\right)^2
=&
\frac{2r_0^3}{r_0-M}
+
\alpha
\left[
24
-
\frac{4(r_0-3M)^2}{5(r_0-M)^2}
\right]
+
\mathcal O(\alpha^2),\\
\left(r_{\rm sh}^{-}\right)^2
=&
\frac{2r_0^3}{r_0-M}
+
\alpha
\left[
-24
-
\frac{4(r_0-3M)^2}{5(r_0-M)^2}
\right]
+
\mathcal O(\alpha^2).
\end{aligned}
\end{equation}
Figure \ref{fig:Shadow} illustrates the variation of the shadow in terms of $q/M$ and $\alpha/M^2$.
This figure indicates that the shadow size is influenced by both the electric charge $q/M$ and the Weyl correction parameter $\alpha/M^2$. When the Weyl parameter is held constant, increasing the electric charge consistently reduces the shadow size. This steady decrease shows that the electromagnetic field's impact on the gravitational potential dominates the overall size of the shadow. However, the behavior becomes more complex when we look at how the Weyl parameter affects different polarization states. We find that the effect of $\alpha/M^2$ on the shadow radius depends on the polarization and completely reverses in trend; for positive polarization, as the Weyl parameter increases from $\alpha/M^2 = -0.025$ to $\alpha/M^2 = 0.025$ while keeping the charge constant, the shadow radius increases; in contrast, for the negative polarization state, increasing $\alpha/M^2$ from $-0.025$ to $0.025$ leads to a decrease in the shadow radius. Accordingly, for $\alpha = 0$, the black hole shadow is identical for both polarization modes, $W_{+}$ and $W_{-}$, and coincides with the shadow of the RN black hole. However, for negative values of $\alpha/M^2$ and fixed $q/M$, the shadow corresponding to the negative polarization mode is larger than the RN shadow, whereas the shadow associated with the positive polarization mode is smaller. Conversely, for positive values of $\alpha/M^2$, this behavior is reversed. This polarization-driven reversal suggests that the Weyl-Maxwell coupling changes the effective refractive index of spacetime differently for each mode. While the charge generally reduces the shadow size, the Weyl parameter either increases or decreases it based on the polarization state of the photon. This unique characteristic could potentially create a birefringence-like effect at the shadow's edge, offering a way to test Weyl-corrected gravity through polarized imaging.

To connect our theoretical model with real astrophysics, we have combined the observational limits from the Event Horizon Telescope (EHT) imaging of Sgr A$^{*}$ with our shadow radius results. By comparing the predicted shadow size to the measured angular diameter of Sgr A$^{*}$, we can set strict limits on the parameters $\alpha$ and $q$. Based on the observational data from Sgr A$^{*}$, the permitted range for the shadow radius is found to be between $4.55$ and $5.22$ at the $1\sigma$ region, while it extends from $4.21$ to $5.56$ at the $2\sigma$ region \cite{vagnozzi2023horizon}. As illustrated in Fig. \ref{fig:Shadow}, our results show that the shadow radius remains entirely within the $2\sigma$ region, while it is more strictly constrained by the $1\sigma$ limit. These findings indicate that only specific regions of the parameter space are physically viable under current EHT observations. These limits show that while the Weyl coupling creates unique birefringence-like effects, its size is tightly constrained by current horizon-scale observations.
\section{Null Geodesics} \label{sec2.1}
Photon propagation in the strong-field region is governed by null geodesics of the effective spacetime. The shadow boundary is determined by the unstable circular photon orbits and the corresponding critical impact parameter. In theories with non-minimally coupled electromagnetic effects, the effective geometry that photons experience can differ from the background spacetime. This can lead to observable changes in the photon sphere and shadow structure~\cite{Fathi2025Jan,Gohain2024Dec,Saikia2025Nov,Haditale2023Sep,Villanueva2013Jun}. To create the optical appearance of the black hole, we use the backward ray tracing method. In this approach, we trace the paths of photons backward from the observer's image plane into the curved spacetime. Depending on the initial conditions, the photons can escape to infinity, intersect the emitting region, or be captured by the black hole event horizon.

	Building upon the effective optical geometry established in Sec. \ref{Sec2}, photon propagation ($\delta^{m} = 0$) in the strong-field region is governed by the null geodesics of metric \eqref{dseff}. Restricting our consideration to the equatorial plane ($\theta = \pi/2$) and utilizing the conserved energy $E$ and angular momentum $L$ introduced via the Lagrangian framework in Sec. \ref{Sec2}, the set of equations of motion governing null trajectories, using the conjugate momentum in $r$-coordinate ($p_r = \partial \mathcal{L}/\partial \dot{r}$), is expressed as
\begin{equation}
	\begin{aligned}
		\dot{t} &= E f(r)^{-1}, \qquad
		\dot{\phi} = \frac{L}{R_{\textrm{eff}}^s(r)}, \qquad
		\dot{r} = p_r f(r),
	\end{aligned}
	\label{geod_eqs}
\end{equation}
with the radial momentum evolution given by
\begin{equation}
	\begin{aligned}
	\dot{p}_r^{s} &=
	\frac{1}{2}\left(- \frac{\partial f(r)}{\partial r}\dot{t}^2 + \frac{\partial f(r)^{-1}}{\partial r}\dot{r}^2 + \frac{\partial R_{\textrm{eff}}^s(r)}{\partial r} \dot{\phi}^2 \right)
	\\&
	= \frac{1}{2}\left(- \frac{E^2}{f(r)^2} \frac{\partial f(r)}{\partial r} + p_r^2 f(r)^2 \frac{\partial f(r)^{-1}}{\partial r} + \frac{L^2 \left.R_{\textrm{eff}}^s\right.'(r)}{R_{\textrm{eff}}^s(r)^2}\right),
	\label{geod_compact}
	\end{aligned}
\end{equation}
where the polarization-dependent effective angular radius $R_{\rm eff}^s(r)$ was introduced in Eq. \eqref{Reff}.

For numerical ray-tracing and analytical clarity, it is instructive to explicitly expand these equations of motion to first order in the Weyl coupling parameter $\alpha$. Employing Eqs. \eqref{Reff2} and \eqref{rhos},
\begin{equation}
	\frac{1}{R_{\rm eff}^{s}(r)}
	=
	\frac{1}{r^2}
	-\alpha \frac{\rho^s(r)}{r^4}
	+\mathcal{O}(\alpha^2),
\end{equation}
and taking $f(r)=f_0(r)+\alpha f_1(r)$, the perturbative expansion of the time component up to first order in $\alpha$ is given by
\begin{equation}
	\dot t
	=
	\frac{E}{f_0(r)}
	-\alpha \frac{E f_1(r)}{f_0^2(r)}
	+\mathcal{O}(\alpha^2).
\end{equation}
Substituting the explicit RN metric component $f_0(r) = 1 - 2M/r + q^2/r^2$ along with the first-order Weyl correction $f_1(r)$, we obtain
\begin{equation}
	\dot t
	=
	\frac{E}{1-\frac{2M}{r}+\frac{q^2}{r^2}}
	+
	\alpha
	\frac{
		E
		\left[
		\frac{4q^2}{3r^4}
		-\frac{40Mq^2}{9r^5}
		+\frac{104q^4}{45r^6}
		\right]
	}
	{
		\left(
		1-\frac{2M}{r}+\frac{q^2}{r^2}
		\right)^2
	}
	+\mathcal{O}(\alpha^2).
\end{equation}
Similarly, the radial equation of motion expands linearly as
\begin{equation}
	\dot r=p_r f(r)
	\Longrightarrow
	\dot r
	=
	p_r f_0(r)
	+
	\alpha p_r f_1(r)
	+
	\mathcal{O}(\alpha^2),
\end{equation}
which, upon inserting the explicit metric functions, yields
\begin{equation}
	\dot r
	=
	p_r
	\left(
	1-\frac{2M}{r}+\frac{q^2}{r^2}
	\right)
	+
	\alpha p_r
	\left(
	-\frac{4q^2}{3r^4}
	+\frac{40Mq^2}{9r^5}
	-\frac{104q^4}{45r^6}
	\right)
	+
	\mathcal{O}(\alpha^2).
\end{equation}
Next, the evolution of the azimuthal angle is determined using Eq. \eqref{geod_compact}. Expanding this relation to linear order in $\alpha$ gives
\begin{equation}
	\dot{\phi^s}
	=
	\frac{L}{R_{\rm eff}^{s}(r)}
	\Longrightarrow
	\dot{\phi^s}
	=
	\frac{L}{r^2}
	-\alpha L\frac{\rho^s(r)}{r^4}
	+\mathcal{O}(\alpha^2).
\end{equation}
For the $W_+$ and $W_-$ polarizations, the above equation gives
\begin{equation}
	\begin{aligned}
\dot\phi^+
=&
\frac{L}{r^2}
+
\alpha L
\left(
-\frac{24M}{r^5}
+\frac{212q^2}{9r^6}
\right)
+
\mathcal{O}(\alpha^2),
\\
\dot\phi^-
=&
\frac{L}{r^2}
+
\alpha L
\left(
\frac{24M}{r^5}
-\frac{220q^2}{9r^6}
\right)
+
\mathcal{O}(\alpha^2).
\end{aligned}
\end{equation}
Finally, the equation governing the radial momentum rate $\dot{p}_r^s$ can be decomposed into the unperturbed background part and the first-order Weyl correction,
\begin{equation}
\dot p_r^{\,s}
=
\dot p_{r,0}
+
\alpha \dot p_{r,1}^{\,\pm}
+
\mathcal{O}(\alpha^2),
\end{equation}
where the ordinary RN part is expressed as
\begin{equation}
\dot p_{r,0}
=
-\frac{E^2}{2}
\frac{f_0'(r)}{f_0^2(r)}
-\frac{p_r^2}{2}f_0'(r)
+\frac{L^2}{r^3},
\end{equation}
and the linear perturbation takes the form
\begin{equation}
\dot p_{r,1}^{s}
=
-\frac{E^2}{2}
\left[
\frac{f_1'(r)}{f_0^2(r)}
-
\frac{2f_1(r)f_0'(r)}{f_0^3(r)}
\right]
-\frac{p_r^2}{2}f_1'(r)
+
\frac{L^2}{2}
\left[
\frac{\left.\rho^{s}\right.'(r)}{r^4}
-
\frac{4\left.\rho^{s}\right.(r)}{r^5}
\right].
\end{equation}
For the $W_+$ polarization, the angular-sector part of the correction becomes
\begin{equation}
\frac{L^2}{2}
\left[
\frac{\left.\rho^{+}\right.'(r)}{r^4}
-
\frac{4\rho^+(r)}{r^5}
\right]
=
L^2
\left(
-\frac{60M}{r^6}
+\frac{212q^2}{3r^7}
\right),
\end{equation}
while for the $W_-$ polarization, it becomes
\begin{equation}
\frac{L^2}{2}
\left[
\frac{\left.\rho^{-}\right.'(r)}{r^4}
-
\frac{4\rho^-(r)}{r^5}
\right]
=
L^2
\left(
\frac{60M}{r^6}
-\frac{220q^2}{3r^7}
\right).
\end{equation}
These expressions reduce to the ordinary RN null geodesic equations when $\alpha\rightarrow 0$.
In practice, since the explicit expanded form of $\dot{p}_r^s$ becomes computationally heavy, it is often preferable for numerical ray tracing to employ the compact form
\begin{equation}
\dot p_r^{s}
=
\frac12
\left[
-\frac{E^2 f'(r)}{f^2(r)}
-p_r^2 f'(r)
+
\frac{L^2\left(R_{\rm eff}^{s}\right)'(r)}
{\left(R_{\rm eff}^{s}(r)\right)^2}
\right].
\end{equation}
Now,
circular photon orbits (constant-$r$) must satisfy $\dot r=0 \implies V_{\mathrm{eff}}(r)=E^{2}$ and $\ddot r=0 \implies V_{\mathrm{eff}}'(r)=0$.
Consequently, the radius $r_{\mathrm{ph}}$ of a circular null orbit is determined by
\begin{equation}
	\frac{\mathrm{d}}{\mathrm{d}r}\left(\frac{f(r)}{R_{\textrm{eff}}^s(r)}\right)\bigg|_{r=r_{\mathrm{ph}}}=0.
\end{equation}
The critical impact parameter $b_{\textrm{c}}=L/E$ for the photon sphere satisfies
\begin{equation}
	b_{\mathrm{c}}^{2} = \left. \frac{R_{\mathrm{eff}}^s(r)}{f(r)} \right|_{r=r_{\mathrm{ph}}},
\end{equation}
null geodesics with $b<b_{\textrm{c}}$ are captured  by the black hole, whereas those with $b>b_{\textrm{c}}$ are scattered to infinity. The stability of the circular null orbit is dictated by the sign of the second derivative of the effective potential: $V_{\textrm{eff}}''(r_{\textrm{ph}})>0$ indicates a stable photon orbit, whereas $V_{\textrm{eff}}''(r_{\textrm{ph}})<0$ corresponds to the standard unstable photon sphere encountered in black hole spacetimes. Thus, the effective potential analysis fully characterizes photon capture, scattering, and the existence (or absence) of photon spheres in this geometry.

The null geodesics are obtained by integrating the equations of motion for the $W_+$ and $W_-$ polarization modes via the backward raytracing technique, where light rays are traced backward from the observer's position into the curved black hole spacetime.

\begin{figure*}[!htb]
	\centerline{\includegraphics[scale=0.27]{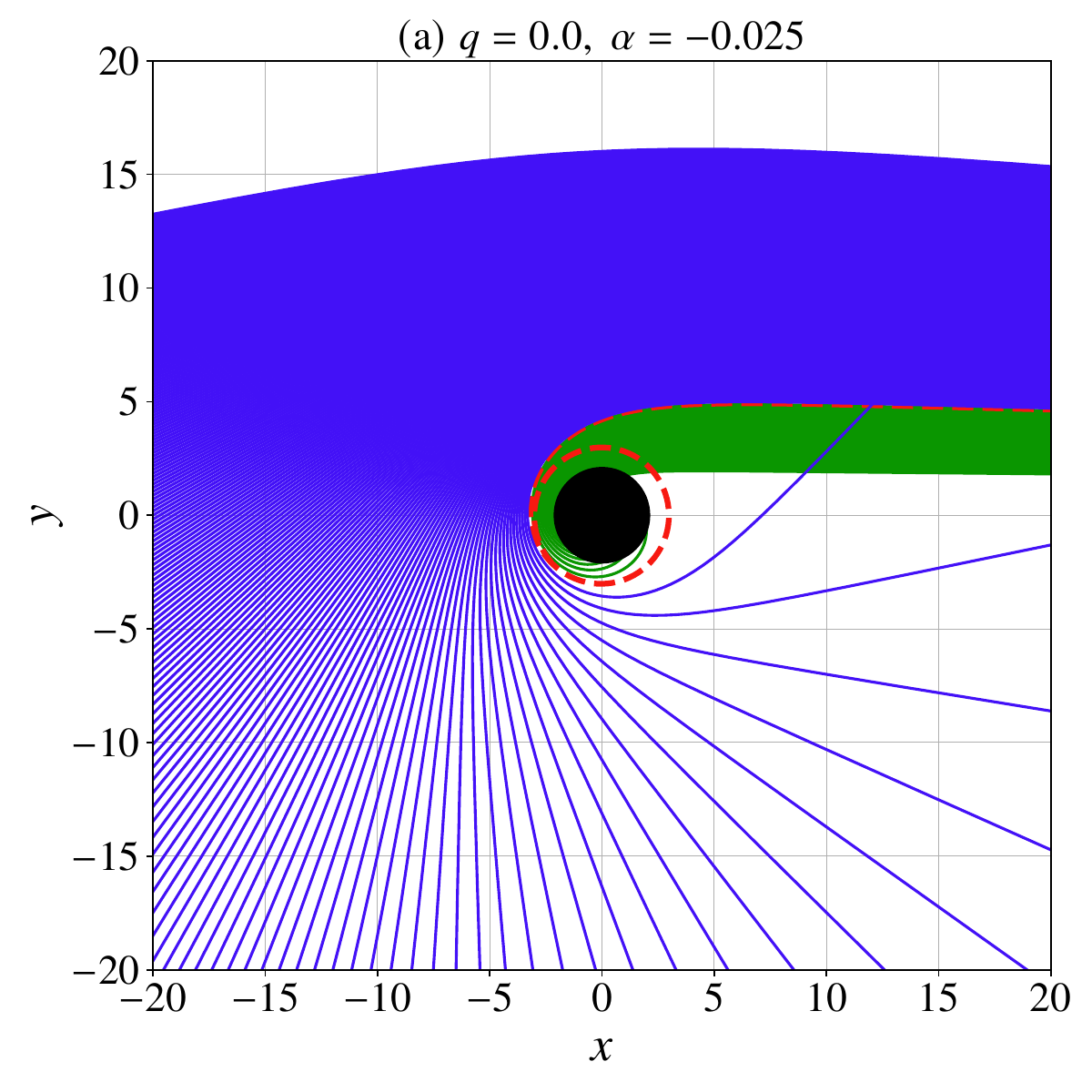} \includegraphics[scale=0.27]{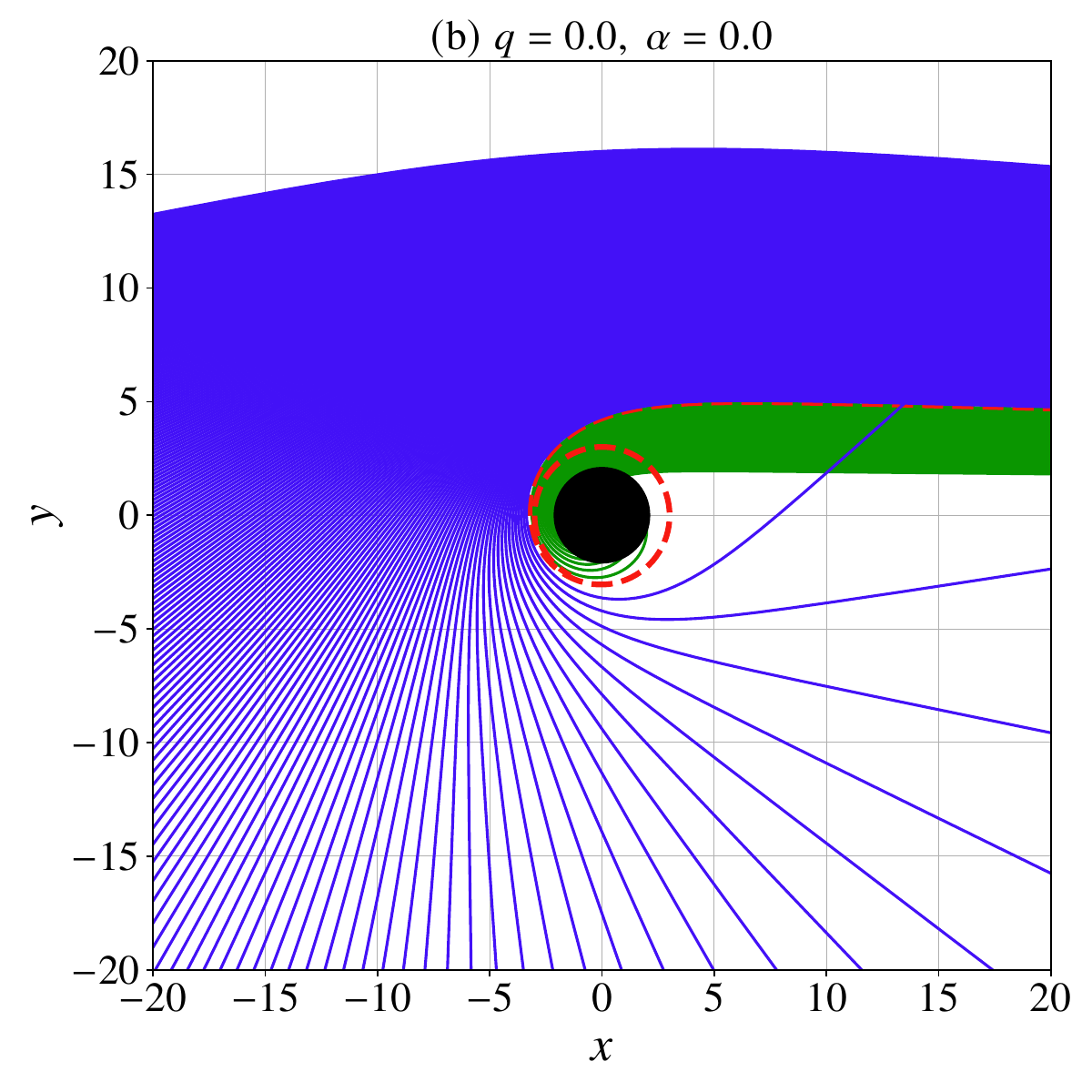} \includegraphics[scale=0.27]{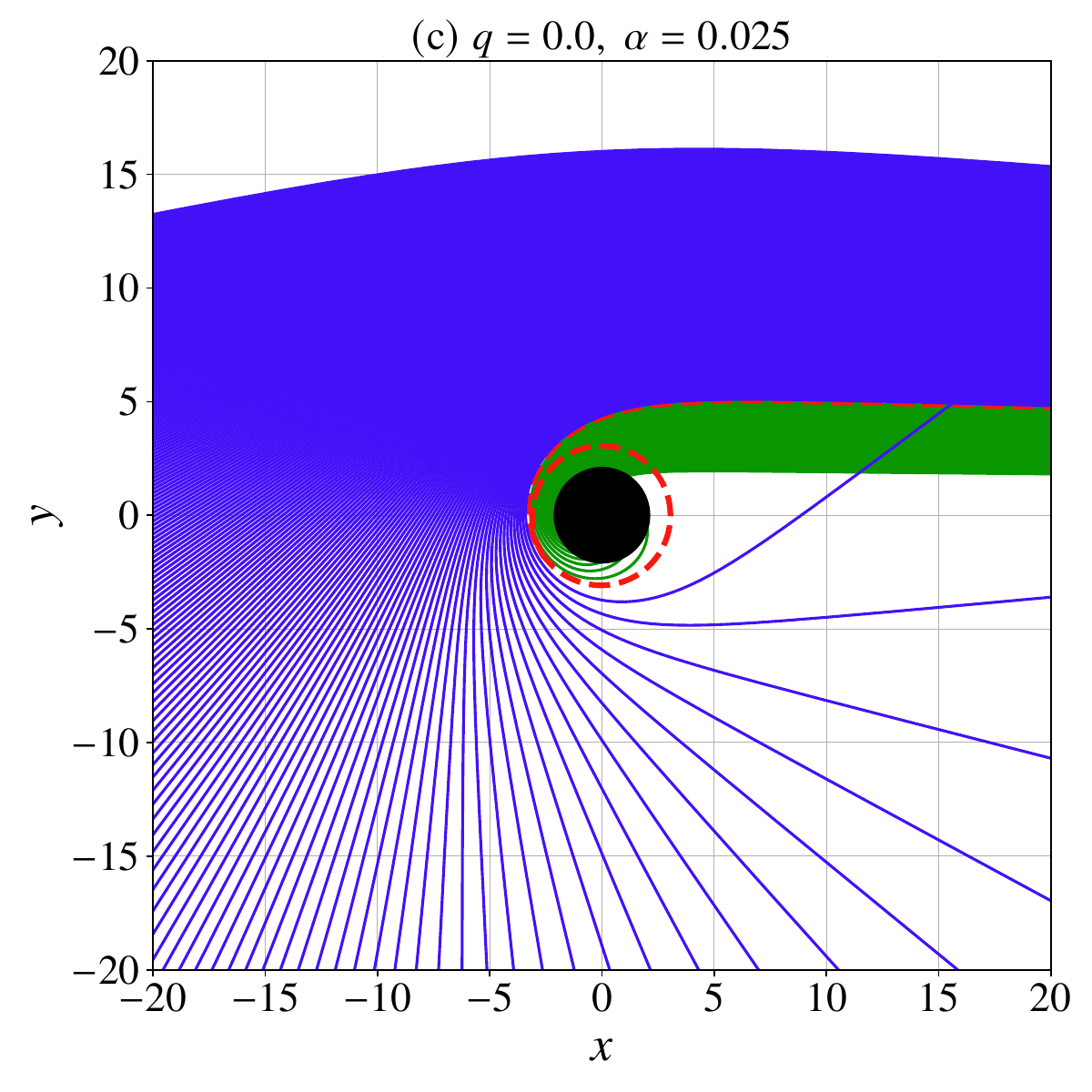}}
	\caption{The null trajectories are shown for different values of $\alpha$ for $W_+$ polarization while keeping $q$ fixed at 0.0.}
	\label{null_fig1} 
\end{figure*}

\begin{figure*}[!htb]
	\centerline{\includegraphics[scale=0.27]{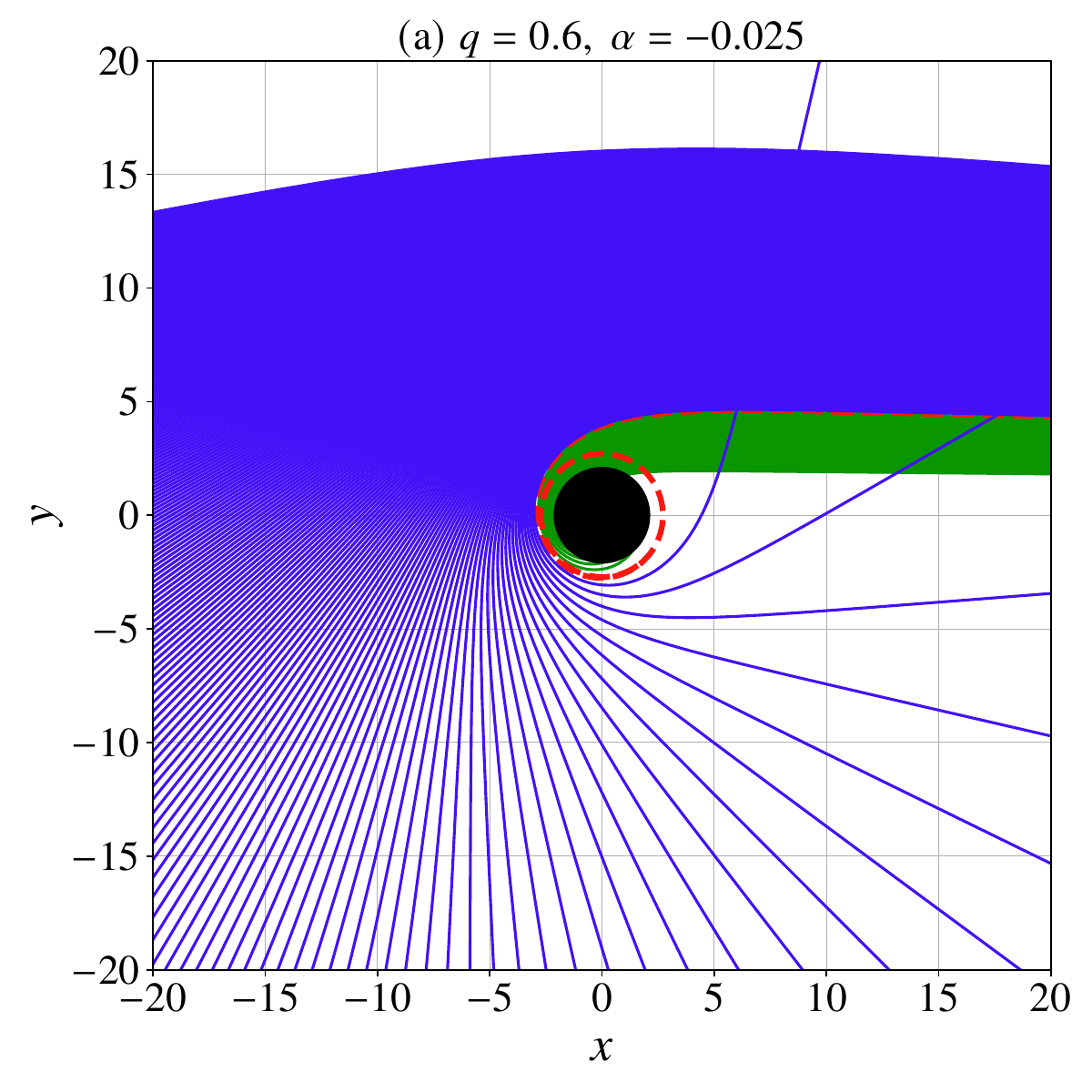} \includegraphics[scale=0.27]{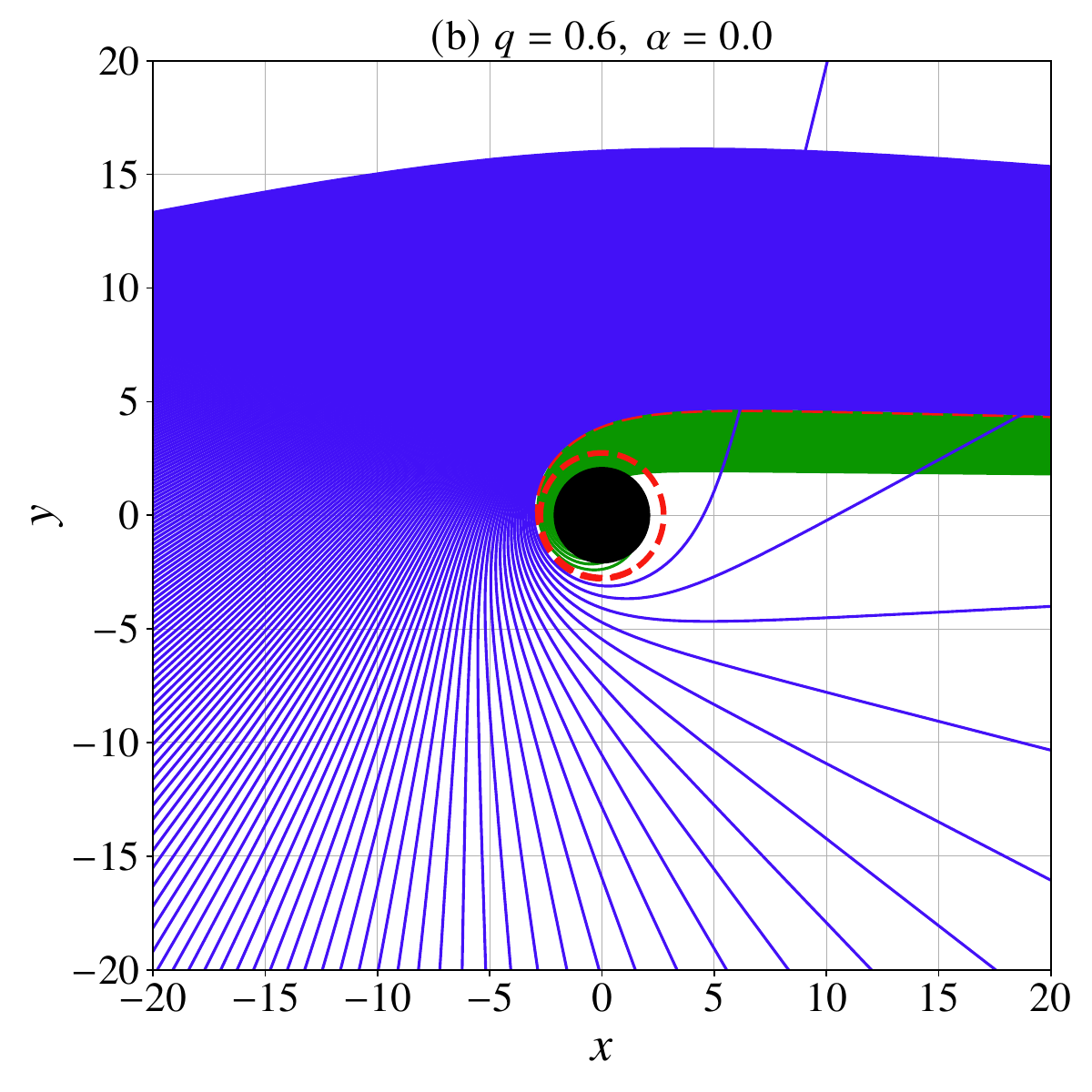} \includegraphics[scale=0.27]{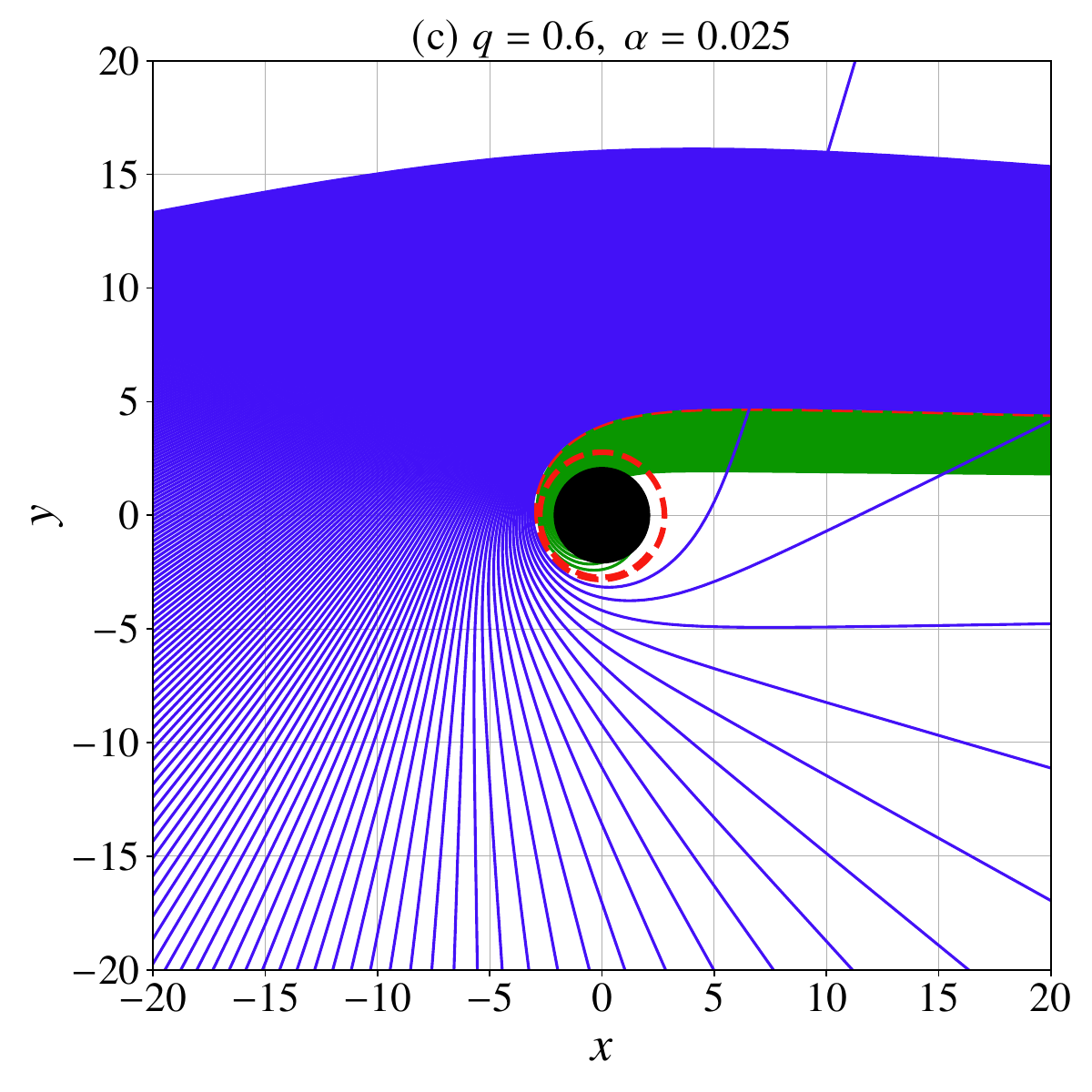}}
	\caption{The null trajectories are shown for different values of $\alpha$ for $W_+$ polarization while keeping $q$ fixed at 0.6.}
	\label{null_fig2} 
\end{figure*}

\begin{figure*}[!htb]
	\centerline{\includegraphics[scale=0.27]{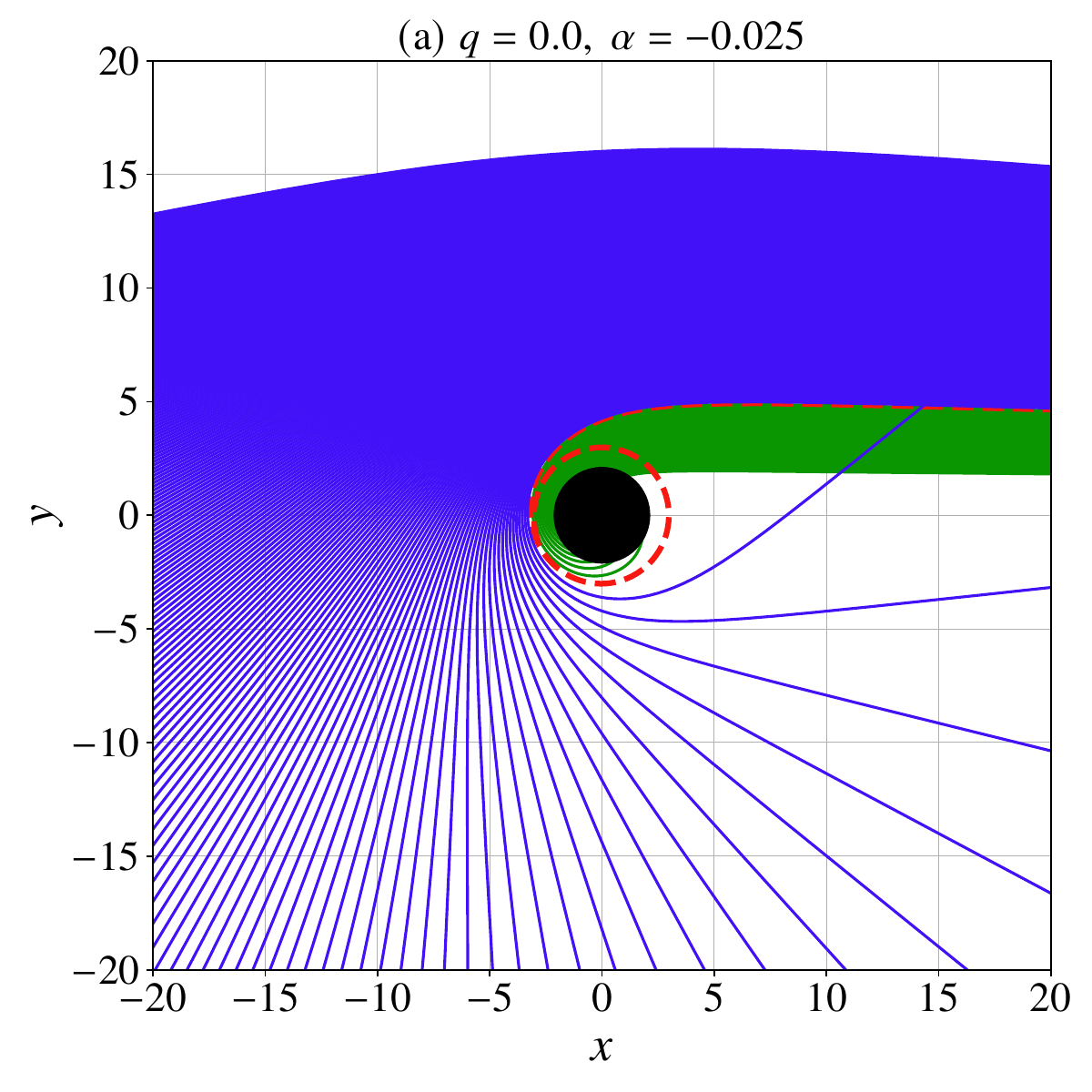} \includegraphics[scale=0.27]{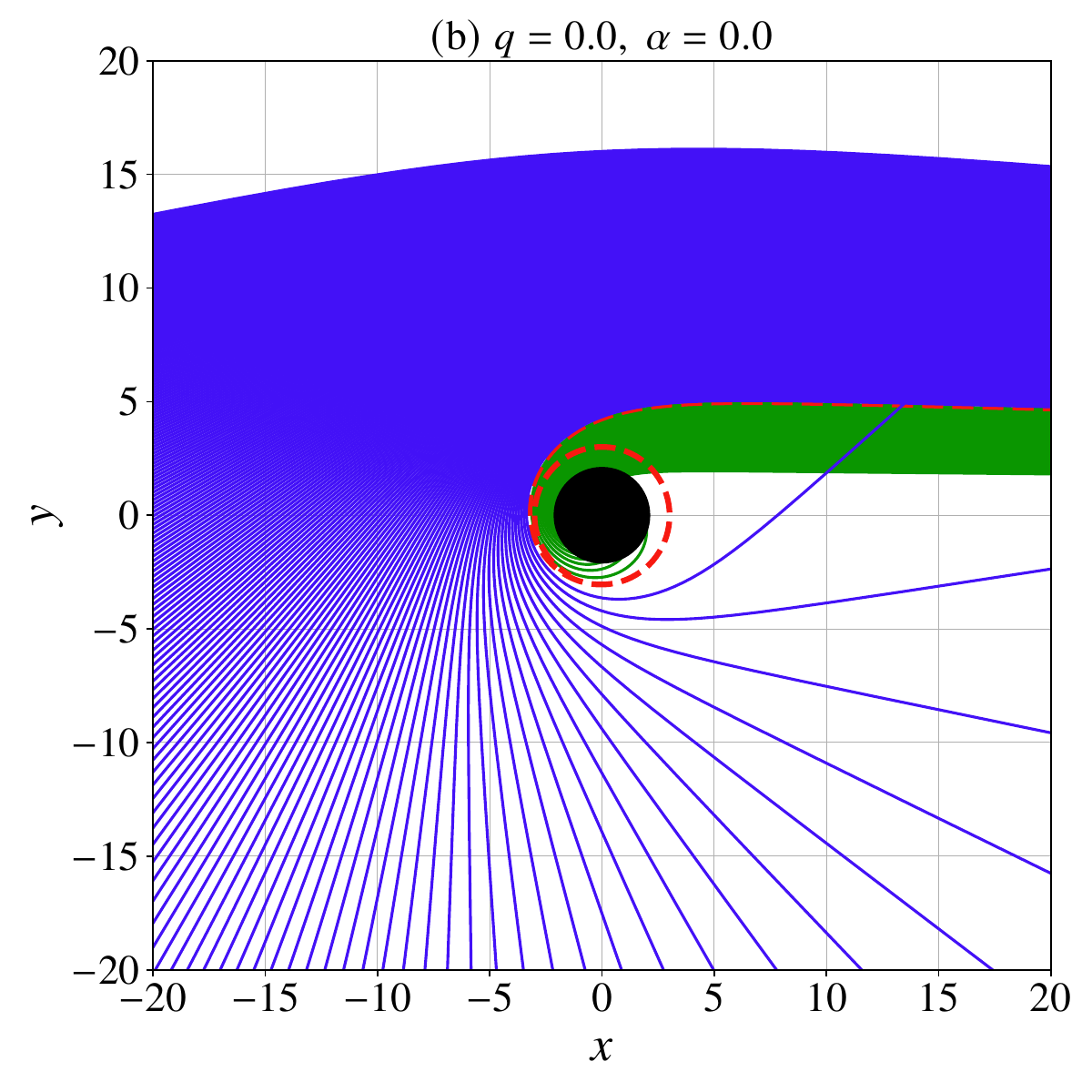} \includegraphics[scale=0.27]{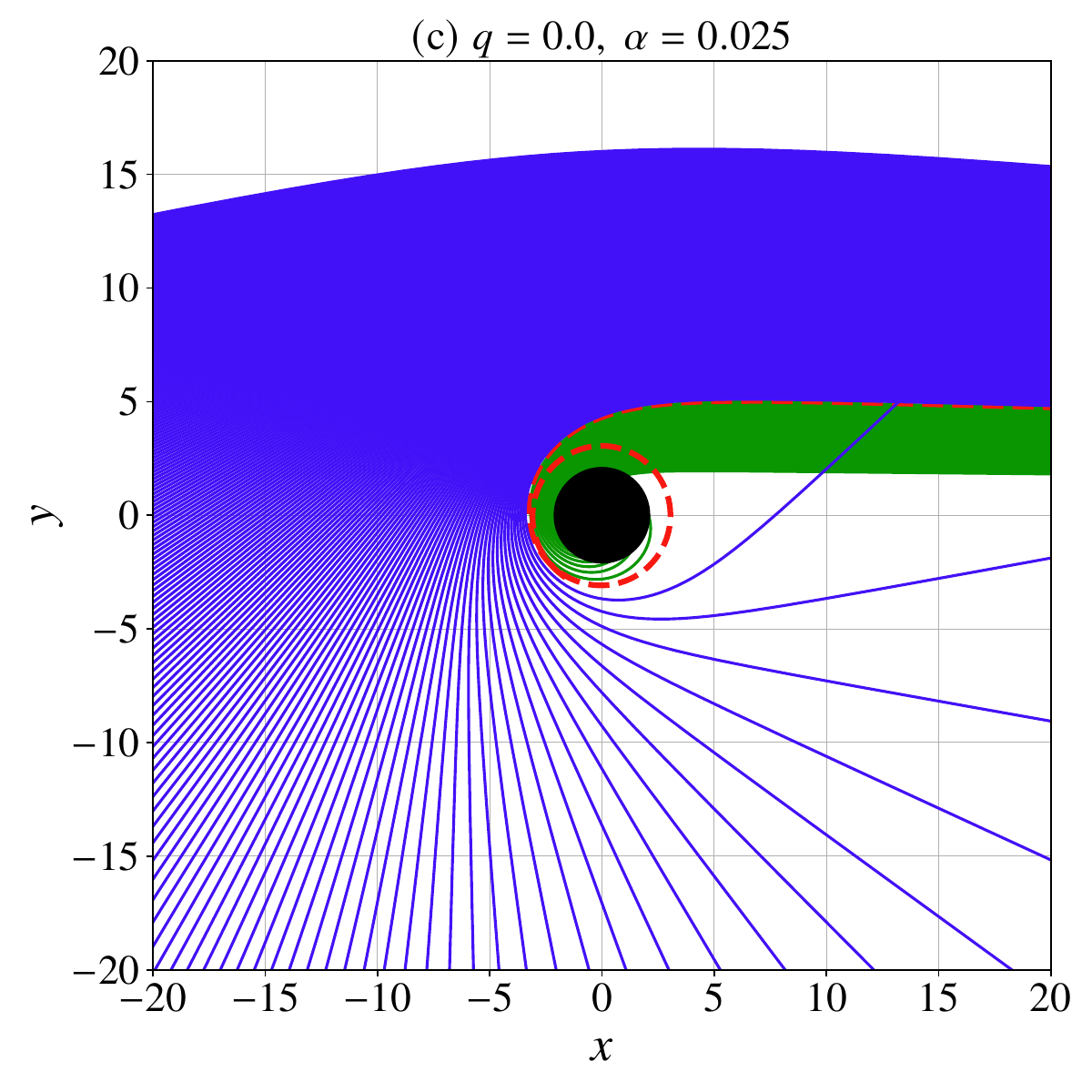}}
	\caption{The null trajectories are shown for different values of $\alpha$ for $W_-$ polarization while keeping $q$ fixed at 0.0.}
	\label{null_fig3} 
\end{figure*}

\begin{figure*}[!htb]
	\centerline{\includegraphics[scale=0.27]{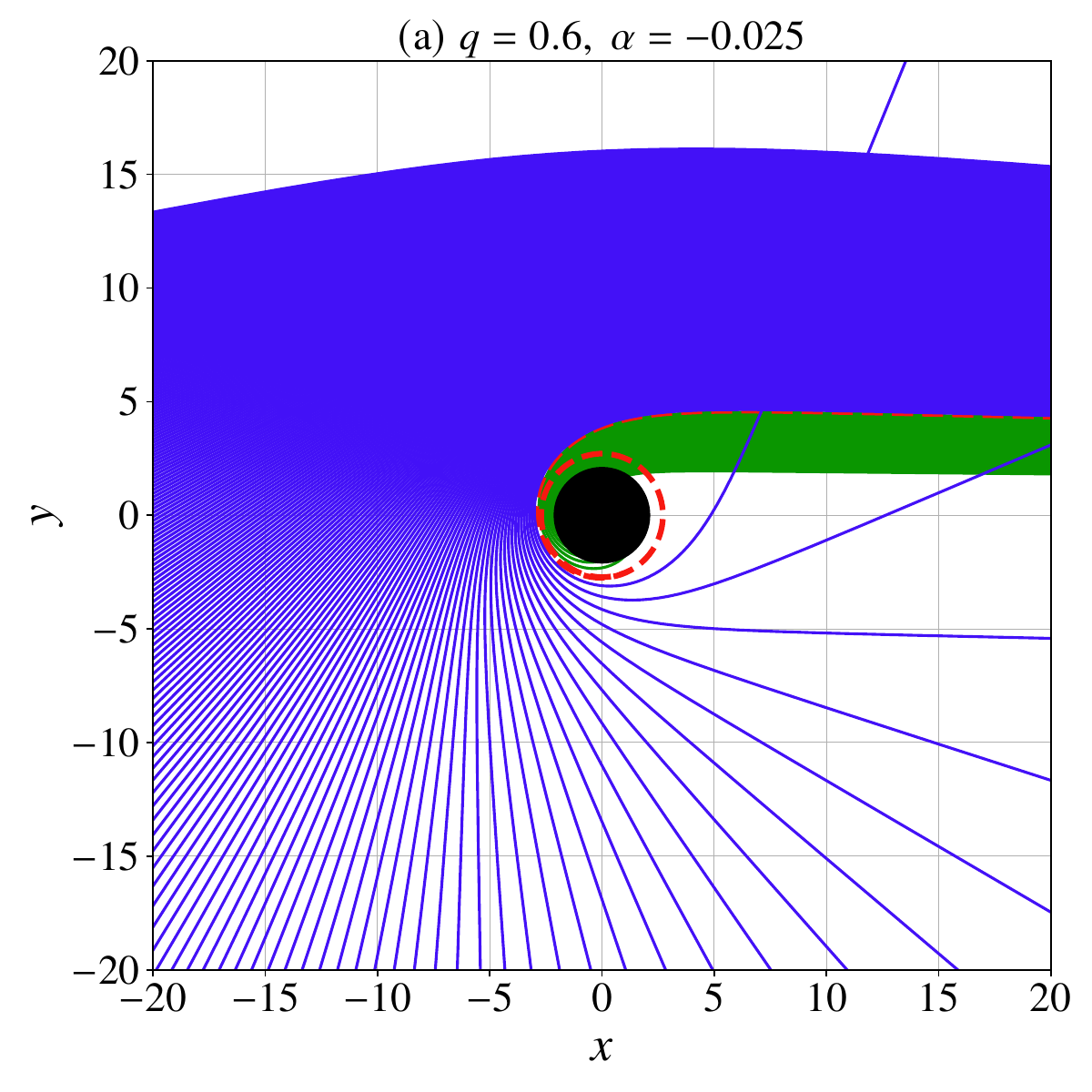} \includegraphics[scale=0.27]{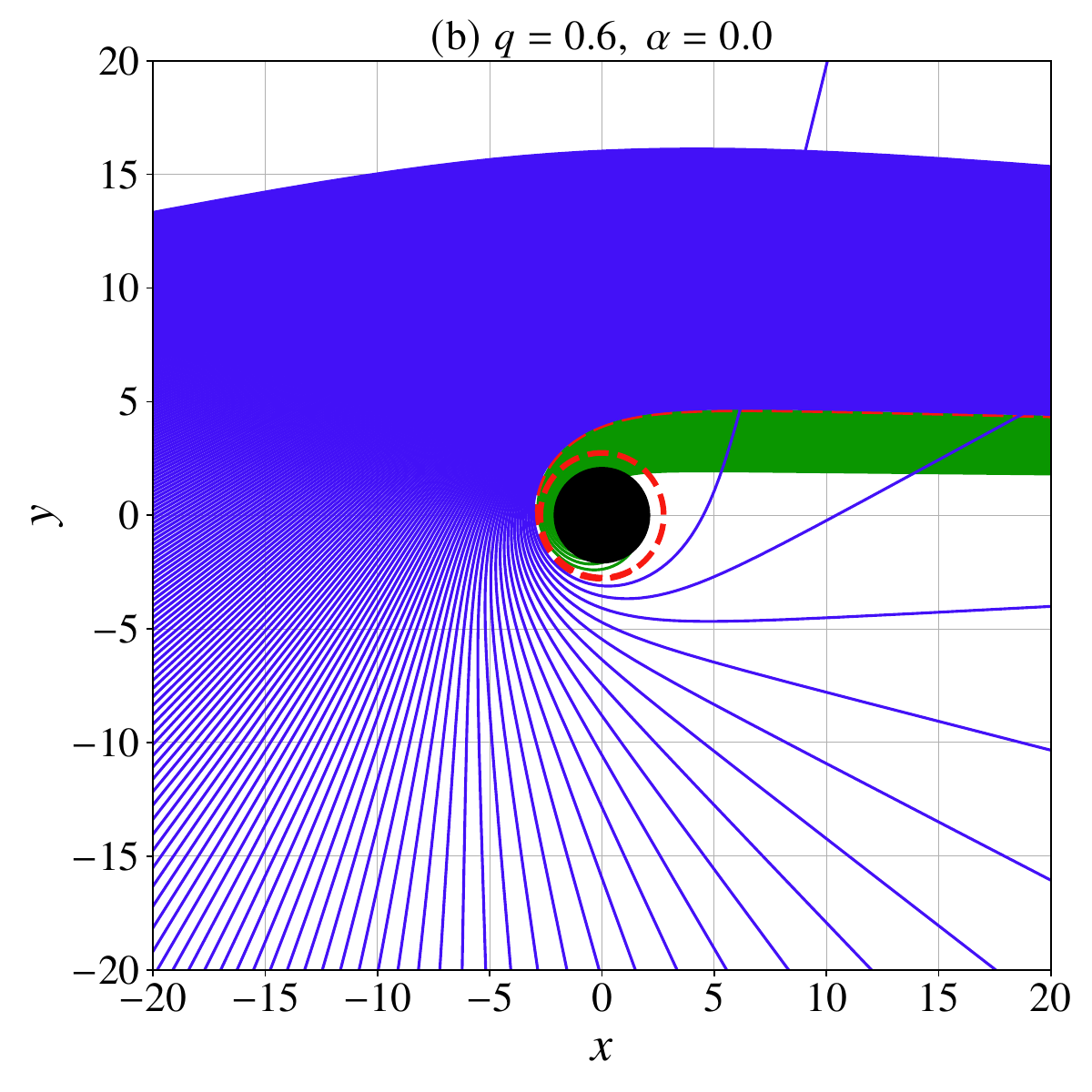} \includegraphics[scale=0.27]{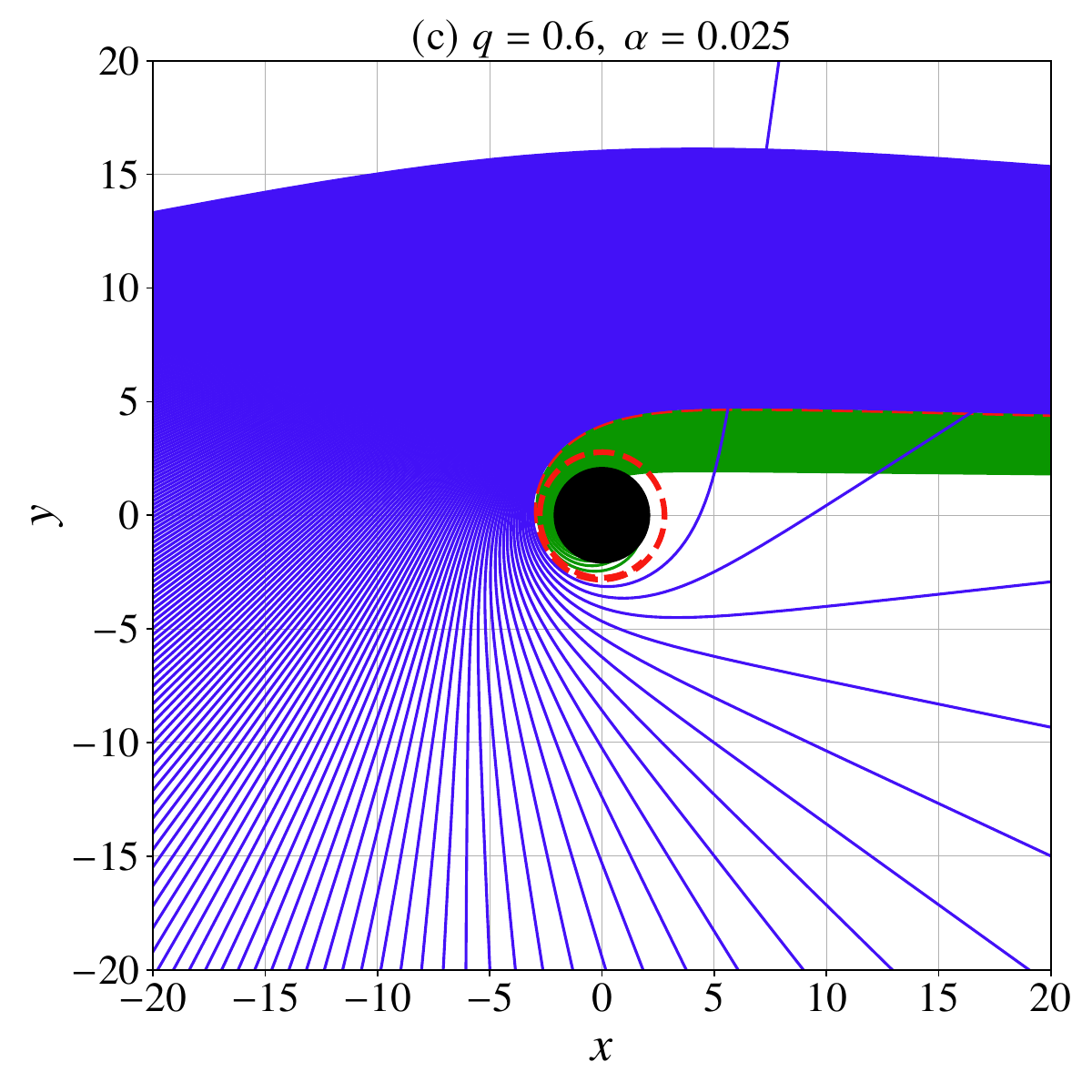}}
	\caption{The null trajectories are shown for different values of $\alpha$ for $W_-$ polarization while keeping $q$ fixed at 0.6.}
	\label{null_fig4} 
\end{figure*}

In our analysis, $E$ is set to 1. The null geodesics are obtained by solving the set of geodesic equations numerically for $W_+$ and $W_-$ polarization modes via the backward raytracing technique, where light rays are traced backward from the observer's position into the curved black hole spacetime. Figures \ref{null_fig1} to \ref{null_fig4} display the resulting trajectories. The circular photon orbits are represented by the red dotted lines.

\begin{table}[ht!]
	\centering
	\begin{tabular}{c|c}
		\hline
		\textbf{$W_+$} & \textbf{$W_-$} \\
		\hline
		\begin{tabular}{ccc}
			\hline
			$\alpha/M^2$ & $q/M$ & $r_{\mathrm{ph}}/M$ \\
			\hline
			$-0.025$ & $0.5$ & $2.78884$ \\
			\hline
			$0.0$  & $0.5$ & $2.82288$ \\
			\hline
			$0.025$  & $0.5$ & $2.85691$ \\
			\hline
		\end{tabular}
		&
		\begin{tabular}{ccc}
			\hline
			$\alpha/M^2$ & $q/M$ & $r_{\mathrm{ph}}/M$ \\
			\hline
			$-0.025$ & $0.5$ & $2.85747$ \\
			\hline
			$0.0$  & $0.5$ & $2.82288$ \\
			\hline
			$0.025$  & $0.5$ & $2.78829$ \\
			\hline
		\end{tabular}
		\\
		\hline
	\end{tabular}
	
	\caption{Photon orbit radius $r_{\mathrm{ph}}$ for $W_{+}$ (left) and $W_{-}$ (right) polarizations.}
	\label{tab:rph}
\end{table}

\noindent
As summarized in Table \ref{tab:rph}, the photon sphere radii exhibit an apparent polarization-antisymmetric dependence on the Weyl coupling parameter at fixed charge $q = 0.5$.  As $\alpha$ varies over the interval $[-0.025,0.025]$, the circular null orbit associated with the $W_{+}$ polarization shifts monotonically outward, whereas the corresponding orbit for the $W_{-}$ polarization shifts inward. The two polarization branches coincide at $\alpha=0$, which recovers the standard (non-birefringent) configuration.

This behaviour reflects the polarization-dependent modification of the photon trajectories in an electromagnetic spacetime induced by the Weyl coupling. In the geometric-optics limit, electromagnetic perturbations propagate along null geodesics of distinct effective metrics, whose potentials differ by terms linear in $\alpha$ with opposite sign for the two polarization states. Consequently, the extremum condition determining the unstable circular orbit, $\mathrm{d}/\mathrm{d}r\,V_{\mathrm{eff}}^{l}(r)=0$, acquires opposite $\alpha$-dependent contributions for $W_{+}$ and $W_{-}$. 
This induces a symmetric birefringent splitting of the photon sphere: an increase in $\alpha$ expands the unstable orbit for one polarization while contracting it for the other, leaving the polarization-averaged mean radius $\left(r_{\mathrm{ph}}^{+}+r_{\mathrm{ph}}^{-}\right)\big/2$ anchored at $r_0$ at linear order in $\alpha$. This antisymmetric structure aligns with the parity-odd nature of polarization-dependent corrections in effective field theory.
\section{Luminosity of Accretion Disks}\label{Sec3}
Accretion disks serve as key astrophysical laboratories for probing strong-field gravitational effects and testing modified black hole geometries. To analyze the thermal radiation profile and particle dynamics in this setting, we employ the standard thin accretion disk model developed by Novikov, Thorne, and Page \cite{novikov1973astrophysics,page1974disk}. In this framework, accreting matter resides near the equatorial plane and moves along circular timelike geodesics. Starting from Eq. \eqref{trajj}, the effective potential for massive test particles is expressed as
\begin{equation}\label{traj}
	V^t_{\rm eff}(r)=\frac{f(r)}{2}\left(1+\frac{L^2}{R(r)}\right),
\end{equation}
which dictates the radial motion via
\begin{equation}
	\frac{1}{2}\dot{r}^2+V^t_{\rm eff}(r)=\frac{1}{2}E^2.
\end{equation}
To study the innermost stable circular orbit, the marginally stable circular trajectory conditions must be satisfied as \cite{jusufi2025black}
\begin{eqnarray}\label{rdot}
	\dot{r}=0\longrightarrow V^t_{\text{eff}}(r)=\frac{1}{2}E^2,\quad \ddot{r}=0\longrightarrow \partial_r V^t_{\text{eff}}(r)=0,\quad\text{and}\quad \dddot{r}=0\longrightarrow\partial^2_r V^t_{\text{eff}}(r)=0.
\end{eqnarray}
We now expand the quantities relevant for circular timelike motion and accretion-disk observables.
The angular velocity of a test particle on the accretion disk is obtained as \cite{jusufi2025black,boshkayev2020accretion,boshkayev2021luminosity}
\begin{equation}\label{omega}
	\Omega(r)=\sqrt{\frac{\partial_r f(r)}{\partial_r R(r)}}.
\end{equation}
The specific angular momentum is calculated fromand specific energy $E(r)$ of the test particle are given by
\begin{equation}
	\label{Lr}
		L(r)=\frac{\Omega(r)R(r)}{\sqrt{f(r)-\Omega^2(r) R(r)}},
\end{equation}
and
\begin{equation}
	\label{Er}
	\begin{aligned}
		E(r)=&\frac{f(r)}{\sqrt{f(r)-\Omega^2(r) R(r)}}.
	\end{aligned}
\end{equation}
Expanding these quantities up to first order in $\alpha$, we obtain the following results.
\\
The angular velocity is expanded as
\begin{equation}\label{Omega1}
\Omega(r)
=
\Omega_{\rm RN}(r)
+
\alpha\Omega_1(r)
+
\mathcal{O}(\alpha^2).
\end{equation}
where the standard RN contribution and the leading Weyl correction read
\begin{equation}
\Omega_{\rm RN}(r)
=
\frac{\sqrt{Mr-q^2}}{r^2}, \qquad
\Omega_1(r)
=
\frac{
	2q^2
	\left(
	30r^2-120Mr+73q^2
	\right)
}
{
	45r^6\sqrt{Mr-q^2}
}.
\end{equation}
In the uncharged limit ($q=0$), Eq. \eqref{omega} reduces to the Schwarzschild angular velocity, $\Omega_{\rm Sch}(r)=\sqrt{M/r^3}$.

Similarly, the specific angular momentum yields the perturbative expansion
\begin{equation}\label{Lr2}
L(r)
=
L_{\rm RN}(r)
+
\alpha L_1(r)
+
\mathcal{O}(\alpha^2),
\end{equation}
where the ordinary RN contribution and the first-order Weyl correction of angular momentum are expressed as
\begin{equation}
L_{\rm RN}(r)
=
\frac{
	r\sqrt{Mr-q^2}
}
{
	\sqrt{r^2-3Mr+2q^2}
},\qquad
L_1(r)
=
\frac{
	2q^2
	\left(
	165M^2r^2
	-150Mq^2r
	-155Mr^3
	+32q^4
	+78q^2r^2
	+30r^4
	\right)
}
{
	45r^3
	\sqrt{Mr-q^2}
	\left(
	r^2-3Mr+2q^2
	\right)^{3/2}
}.
\end{equation}
Setting $q=0$ reproduces the Schwarzschild expression $L_{\rm Sch}(r)
=
\sqrt{Mr}/\sqrt{1-3M/r}$.

Finally, the specific energy is expanded as
\begin{equation}\label{ErW}
E(r)
=
E_{\rm RN}(r)
+
\alpha E_1(r)
+
\mathcal{O}(\alpha^2).
\end{equation}
where the zeroth- and first-order terms are given by
\begin{equation}
E_{\rm RN}(r)
=
\frac{
	r^2-2Mr+q^2
}
{
	r\sqrt{r^2-3Mr+2q^2}
},\qquad
E_1(r)
=
\frac{
	2q^2
	\left(
	30M^2r^2
	+3Mq^2r
	-65Mr^3
	-10q^4
	+27q^2r^2
	+15r^4
	\right)
}
{
	45r^5
	\left(
	r^2-3Mr+2q^2
	\right)^{3/2}
}.
\end{equation}
For \(q=0\), the Schwarzschild limit is recovered
$
E_{\rm Sch}(r)
=
\left(1-2M/r\right)/\left(
\sqrt{1-3M/r}\right).
$

These expressions satisfy two important consistency checks. First, in the limit $\alpha\rightarrow 0$, all quantities reduce to their RN forms. Second, in the neutral limit $q\rightarrow 0$, the Schwarzschild expressions are recovered. Therefore, the first-order expansion consistently separates the ordinary black hole contributions from the leading Weyl corrections.

\begin{figure}[ht!]
		\centerline{
		\includegraphics[width=7cm]{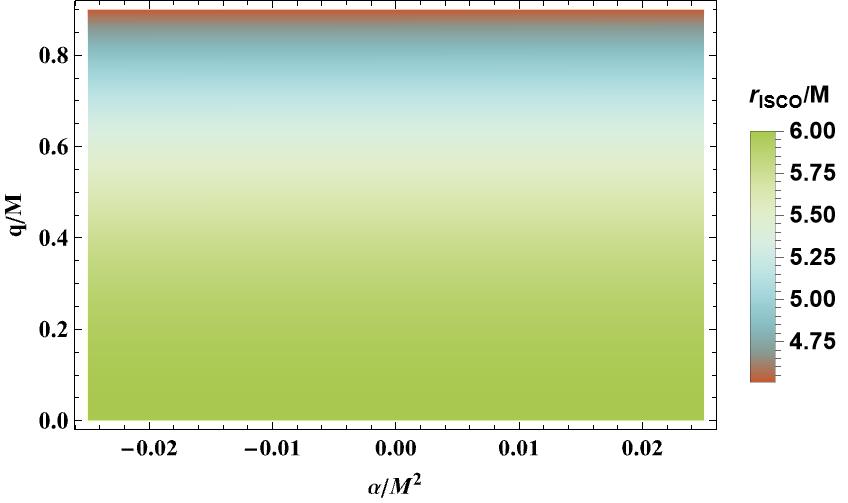}}
	\caption{Black hole ISCO radius as a function of the electric charge and the Weyl correction parameter. The colorbar denotes the variation in the magnitude of the ISCO radius across the parameter space.}\label{fig:ISCO}
\end{figure}

The ISCO radius is determined by evaluating the extremum condition of the specific energy \eqref{rdot} and \eqref{ErW}
\begin{equation}\label{ISCO}
	\partial_r E(r)\Big|_{r=r_{\rm ISCO}}=0.
\end{equation}
The behavior of the ISCO radius as a function of the parameters $\alpha/M^2$ and $q/M$ is demonstrated in Figure \ref{fig:ISCO}.
Our numerical results indicate that for a fixed Weyl parameter, an increase in the electric charge leads to a reduction in the ISCO radius. This suggests that the charge enhances the gravitational binding, allowing stable orbits to exist closer to the horizon. Conversely, shifting the Weyl parameter from $\alpha/M^2 = -0.025$ to $0.025$ results in an increase in the ISCO radius. Although the magnitude of this shift is minimal, it indicates that a positive Weyl coupling introduces a repulsive-like correction to the effective potential, pushing the stable disk boundary outward.

\begin{figure}[ht!]
		\centerline{
		\includegraphics[width=6cm]{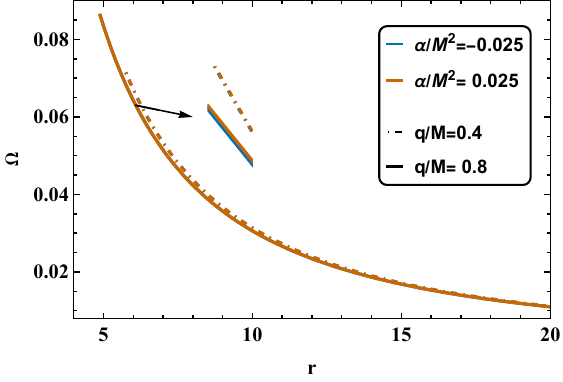} \hspace{0.5cm}
		\includegraphics[width=6cm]{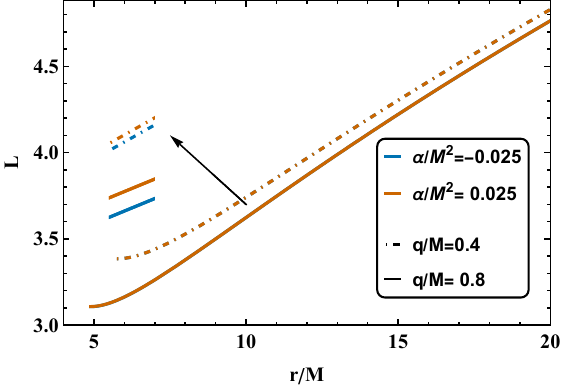} \hspace{-0.2cm}}
		\centerline{
		\includegraphics[width=6cm]{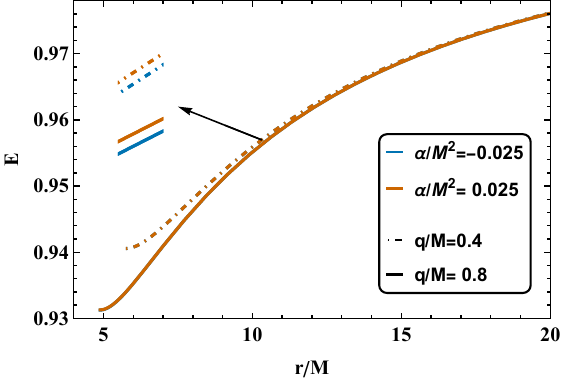} \hspace{0.5cm}
		\includegraphics[width=6cm]{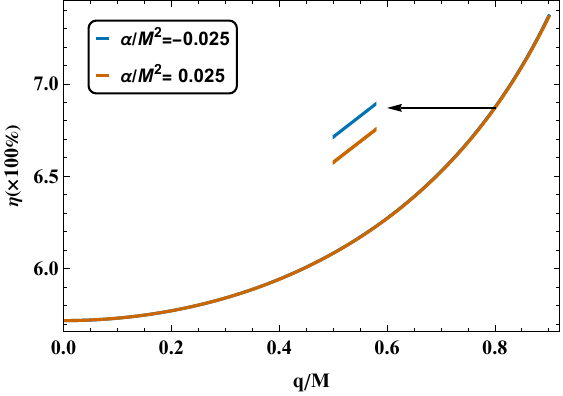}}
	\caption{Accretion disk orbital quantities and radiative efficiency. Upper left panel: angular velocity. Upper right panel: specific angular momentum. Lower left panel: specific energy. Lower right panel: radiative efficiency as a function of $q/M$ for various values of $\alpha/M^2$.}\label{fig:OmegaLE}
\end{figure}

A crucial observable in accretion disk physics is the radiative efficiency which quantifies the conversion of infalling rest mass into radiation. Utilizing Eqs.~\eqref{ErW} and \eqref{ISCO}, the radiative efficiency is expressed as \cite{boshkayev2024luminosity}
\begin{eqnarray}
	\eta=1-E(r_{\rm ISCO}).
\end{eqnarray}
In Figure \ref{fig:OmegaLE}, the radial profiles of angular velocity, angular momentum, specific energy, alongside the variation of the radiative efficiency with respect to the parameter $q/M$ are illustrated.
Both energy and angular momentum exhibit a monotonic growth with radial distance, and energy asymptotically approaches unity at large scales. At a fixed $\alpha/M^2$, increasing the charge parameter leads to a decrease in the energy and angular momentum of the orbiting particles. This effect is most pronounced in the strong-field region (small $r/M$), where the intensified gravitational potential requires less orbital energy for particles to maintain equilibrium. In contrast, increasing $\alpha/M^2$ from $-0.025$ to $0.025$ causes a slight increase in both $E$ and $L$, consistent with the expansion of the ISCO.
The angular velocity follows a declining trend as $r/M$ increases. We found that higher values of $q/M$ lead to a decrease in $\Omega$ at a fixed radius. However, increasing the Weyl parameter results in a slight boost to the angular velocity.
Furthermore, increasing the charge significantly enhances the efficiency, as the disk extends into deeper regions of the potential well, and shifting $\alpha$ from $-0.025$ to $0.025$ results in a reduction of the efficiency. Consequently, the negative Weyl regime combined with a high electric charge represents the most energetically efficient configuration for the accretion process.

\begin{figure}[ht!]
		\centerline{
		\includegraphics[width=6cm]{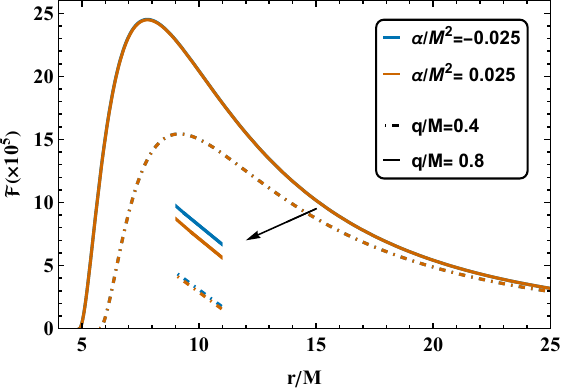} \hspace{0.5cm}
		\includegraphics[width=6cm]{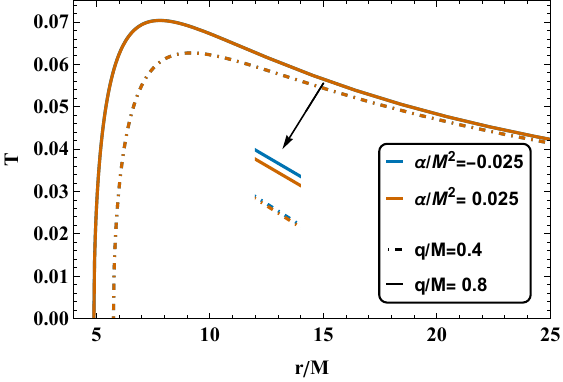}}
	\caption{Radial profiles of the energy flux and radiation temperature of the accretion disk as a function of $r/M$ for varying values of $q/M$ and $\alpha/M^2$.}\label{fig:FT}
\end{figure}

The energy flux of the accretion disk, which represents the radiative flux emitted by the accretion disk, is obtained from
\begin{eqnarray}\label{flux}
	\mathfrak{F}(r)=-\frac{\dot{m}}{4\pi\sqrt{R(r)}}\frac{\partial_r \Omega(r)}{(E(r)-\Omega(r) L(r))^2}\int_{r_{\text{ISCO}}}^{r}(E(\breve{r})-\Omega(\breve{r}) L(\breve{r}))\partial_{\breve{r}}L(\breve{r}) d \breve{r},
\end{eqnarray}
Using the Stefan-Boltzmann law, the effective radiation temperature can be calculated using $\mathfrak{F}(r)=\sigma_{\rm SB}T^4(r)$. In this relation, $\sigma_{\rm SB}$ represents the Stefan-Boltzmann constant.
\\In Figure \ref{fig:FT}, the energy flux and the radiation temperature are displayed for different values of $q/M$ and $\alpha/M^2$.
The energy flux vanishes at the ISCO radius, indicating the inner edge of the radiative disk. Beyond this point, the flux rises sharply to reach a maximum value before exhibiting a slow decay at larger distances. The influence of the model parameters is distinct: at a constant Weyl coupling, an increase in the electric charge leads to a significantly higher peak flux, which occurs at a smaller radial distance. This is consistent with the inward migration of the ISCO, allowing for a more compact and energetic inner disk. Conversely, at a fixed charge, increasing parameter $\alpha$ from $-0.025$ to $0.025$ results in a decrease in the peak flux, and this maximum is shifted toward a larger radius. This indicates that a positive Weyl correction tends to reduce the overall energy output of the disk.
Since the radiation temperature is directly linked to the energy flux via the Stefan--Boltzmann relation, it follows an entirely analogous trend.

\begin{figure}[ht!]
		\centerline{
		\includegraphics[width=7cm]{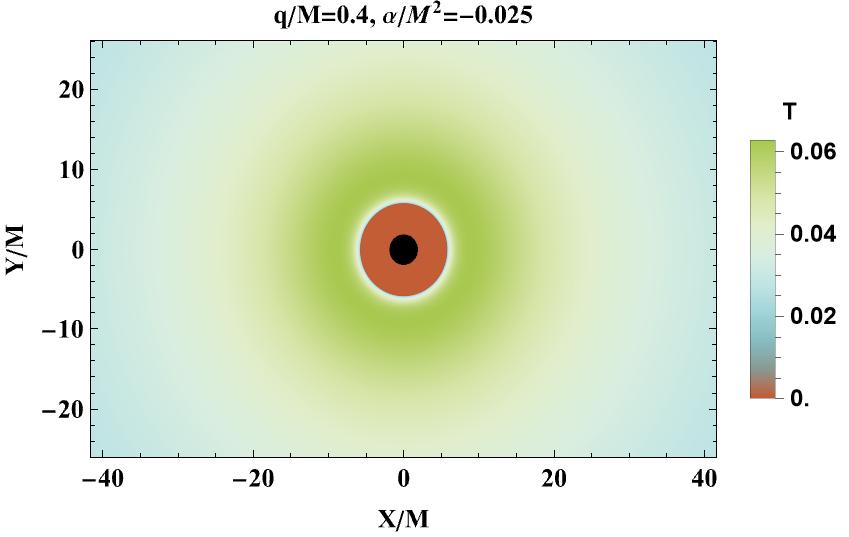} \hspace{0.5cm}
		\includegraphics[width=7cm]{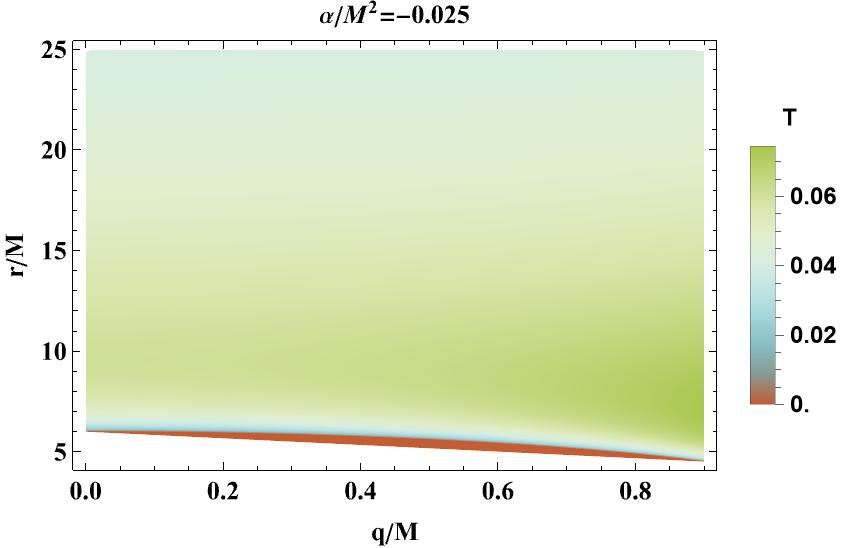}}
	\caption{Radiation temperature of the accretion disk. Left panel: two-dimensional radiation temperature distribution in the $X-Y$ plane for fixed values of $\alpha/M^2$ and $q/M$. Right panel: radiation temperature distribution for $\alpha/M^2=-0.025$ as a function of $r/M$ and $q/M$.}\label{fig:TDen}
\end{figure}

In Figure \ref{fig:TDen}, the radial variations of the radiation temperature are depicted for different values of $q/M$, alongside two-dimensional temperature distributions in the equatorial $X-Y$ plane.
The trend in the changes of the radiation temperature and the impact of the parameter $q/M$ on its magnitude are evident. Increasing the parameter $q/M$ raises the radiation temperature and moves the maximum value to a smaller radius.

\begin{figure}[ht!]
		\centerline{
		\includegraphics[width=6cm]{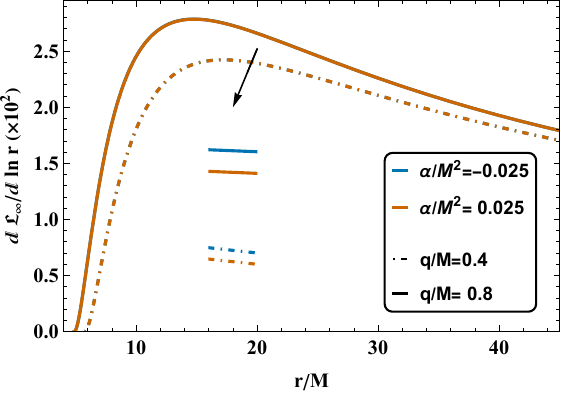} \hspace{0.5cm}
		\includegraphics[width=6cm]{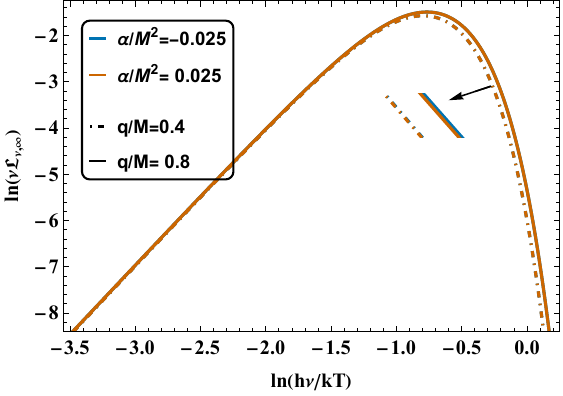}}
	\caption{Differential luminosity (left) and spectral luminosity (right) as a function of $r/M$ and emission frequency, respectively, for varying parameter choices of $q/M$ and $\alpha/M^2$.}\label{fig:Lum}	
\end{figure}

The differential luminosity measure describes the change in luminosity at varying distances from the disk and is obtained from
\begin{eqnarray}
	\frac{d \mathfrak{L}_\infty}{d \ln r}=4\pi r \sqrt{R(r)} E(r) \mathfrak{F}(r),
\end{eqnarray}
Spectral luminosity indicates the amount of energy emitted at different frequencies, illustrating how the disk produces radiation across various frequencies $\nu$, and is given by \cite{boshkayev2024luminosity}
\begin{equation}
	\nu\mathfrak{L}_{\nu,\infty}=\frac{15}{\pi^4}\int_{r_i}^\infty \left(\frac{d \mathfrak{L}_\infty}{d \ln r}\right)\frac{\left(u^t(r)y\right)^4}{M^2 \mathfrak{F}(r)}\frac{1}{\exp\left(u^t(r)y/\left[M^2 \mathfrak{F}(r)\right]^{1/4}\right)-1}d\ln r
	,\qquad u^t(r)=1/\sqrt{f(r)-\Omega^2(r)R(r)},
\end{equation}
where $y=\hbar\nu/k T$ and $\hbar$ represents Planck's constant.
In Figure \ref{fig:Lum}, the variations of differential and spectral luminosity are illustrated.
The differential luminosity follows a profile similar to the energy flux, characterized by an initial peak followed by a gradual decline. Our calculations show that for a constant $\alpha/M^2$, a higher charge shifts the peak of the differential luminosity toward a smaller radius and increases its magnitude. In contrast, increasing the Weyl parameter from $-0.025$ to $0.025$ (at fixed $q/M$) decreases the peak luminosity and moves it to a larger radial distance.
We observed that as $q/M$ increases (at fixed $\alpha/M^2$), the maximum value of the spectral luminosity increases, and the peak shifts toward higher frequencies. This spectral blue-shift is a direct consequence of the disk becoming hotter. However, at a fixed charge, increasing the Weyl parameter $\alpha/M^2$ from $-0.025$ to $0.025$ leads to a reduction in the peak spectral luminosity and shifts it toward lower frequencies.
\hspace{8.5cm}
\section{Conclusion}\label{Sec10}
In this study, we examined how Weyl corrections affect the physical and thermodynamic properties of a charged black hole. By looking at the non-minimal coupling between the Weyl tensor and the Maxwell field, we showed that this model leads to important deviations from the classical Einstein-Maxwell predictions.

Regarding the geometric structure, our analysis of the lapse function and the event horizon reveals that both the electric charge and the Weyl coupling parameter contribute to the contraction of the event horizon radius. While the electric charge remains the dominant factor in determining the black hole's scale, the Weyl parameter acts as a secondary but critical modifier. Specifically, an increase in $\alpha$ further reduces the horizon size, suggesting that the Weyl coupling reinforces the effective repulsive potential in the strong-field regime.

In the thermodynamic section, our results reveal a non-trivial interplay between charge and the Weyl correction parameter. Specifically, the thermodynamic response of the black hole to the charge is strictly governed by the sign of $\alpha$. In the regime of $\alpha > 0$, the Weyl correction exerts a suppressive influence, where increasing the charge leads to a reduction in entropy, suggesting a geometric constraint imposed by the positive correction term. Conversely, for $\alpha < 0$, the correction enhances the effect of charge, causing the entropy to increase monotonically with $q$. We also found that increasing $q$ and $\alpha$ causes a synchronized shift in the Hawking temperature and heat capacity. The divergence points, which signal a phase transition, migrate toward larger radii as these parameters increase.
Crucially, our analysis of the extremal radius shows that it grows monotonically with $q$. This indicates that higher charges provide a stronger barrier against gravitational collapse.
We found that $r_{\rm e}$ is an increasing function of $\alpha$ in the range we studied. As $\alpha$ changes from $-0.025$ to $0.025$, the extremal radius undergoes an expansion. This reveals that a positive Weyl coupling results in a larger extremal radius, while a negative coupling leads to a more compact final state.

In addition, our investigation into the thermodynamic topology via the generalized off-shell free energy provides a global perspective on the system's stability. We found that despite the bifurcations observed in the optical sector, the number of zero points remains invariant for different choices of $\alpha$. This topological constancy confirms that the black hole consistently belongs to the $W^{0+}$ class, indicating that the global thermodynamic phase structure and the arrangement of stable/unstable branches are inherent features of the model, robust against variations in the Weyl coupling.

The shadow size is governed by the interplay between the electric charge and the Weyl correction parameter. For a fixed Weyl parameter, increasing the electric charge always reduces the shadow radius, confirming the dominant role of the electromagnetic field in shrinking the shadow. In contrast, the Weyl correction exhibits a polarization-dependent behavior: at fixed charge, increasing $\alpha/M^2$ enlarges the shadow for the positive polarization state but reduces it for the negative polarization state, revealing an opposite response of the two photon modes. These results demonstrate that the Weyl–Maxwell coupling introduces distinctive polarization-dependent signatures that could provide an observational probe of Weyl-corrected gravity through polarized black hole shadow measurements. To confirm the validity of these findings, we imposed observational limits from Sgr A$^{*}$. By aligning our model with the EHT data, we defined the acceptable parameter space at both $1\sigma$ and $2\sigma$ confidence levels. These limits indicate that while the Weyl coupling permits new features such as polarization-dependent birefringence and shadow growth, its size is strictly constrained by existing horizon-scale observations.

This behavior is further reflected in the accretion disk phenomenology.
We observed that an increase in the electric charge significantly reduces the ISCO radius, thereby intensifying the energy flux and inducing a spectral blue-shift in the luminosity peak toward higher frequencies. In contrast, a positive Weyl parameter $\alpha$ acts as a geometric moderator; it increases the ISCO radius and shifts the spectral peak toward lower frequencies, effectively regulating the thermal intensity of the disk. 

Finally, the null geodesic analysis shows that the effective spacetime felt by photons is strongly influenced by the polarization modes and the Weyl coupling parameter $\alpha$. This leads to different changes in the effective potential, the structure of the photon sphere, and the critical impact parameter. The backward ray-tracing results indicate that light rays propagation and capture depend closely on the polarization states $W_+$ and $W_-$. The Weyl coupling parameter $\alpha$ is important for determining the radius of the photon orbit. Changing $\alpha$ shifts the position of the circular null orbit, which alters the size of the photon sphere. Thus, the coupling parameter directly affects the optical scale of the black hole shadow by influencing the null geodesics. This suggests there could be observable signs of Weyl modifications in black hole optical phenomena.
\hspace{4cm}
\section*{Acknowledgements}
The authors would like to thank the esteemed reviewer for his/her constructive comments. Also,
H. H. is grateful to Excellence project FoS UHK 2203/2025-2026 for the financial support.

\bibliographystyle{ieeetr}
\bibliography{ref.bib}

\end{document}